\documentclass[aps,prb,groupedaddress,twocolumn]{revtex4-2}

 \usepackage{graphicx,bm,amssymb}

\newcommand{\bea}{\begin{eqnarray}}
\newcommand{\eea}{\end{eqnarray}}
\newcommand{\be}{\begin{equation}}
\newcommand{\ee}{\end{equation}}
\newcommand{\eps}{\varepsilon}

\begin{document}

\title{Polarization rotation by external electric field in two-dimensional antiferroelectric squaric acid, H$_2$C$_4$O$_4$. }

\author{A.P. Moina}
\email[]{alla@icmp.lviv.ua}
\affiliation{Institute for Condensed Matter Physics, 1 Svientsitskii St., 79011, Lviv, Ukraine}

\date{\today}

\begin{abstract}
A pseudospin model for description of the influence of the electric field, confined to the plane of sublattice polarization, on the two-dimensional squaric acid antiferroelectrics is developed.
The system behavior is analyzed in terms of the parameters of ferroelectric and antiferroelectric ordering, as well as the non-collinearity angle $\theta$, which is the angle between sublattice polarizations. 
The temperature-electric field $T-E_1$ phase diagram is constructed. Most of the field-induced transitions are found to be associated with polarization rotation. The observed phases include the  collinear ferroelectric and non-collinear ferrielectric with almost antiparallel and perpendicular polarizations of the sublattices, respectively. The diagram also contains two regions, where the non-collinearity angle varies continuously between, nominally, 0 and 90$^\circ$ and between 90$^\circ$ and 180$^\circ$. The first and second order transition lines, supercritical lines, critical end points, and critical points on the $T-E_1$ phase diagram are detected.
\end{abstract}


\maketitle


\section{Introduction}

The squaric acid, H$_2$C$_4$O$_4$ (3,4-dihydroxy-3-cyclobutene-1,2-dione) is a classical example of two-dimensional antiferroelectrics. In these crystals the hydrogen bonded C$_4$O$_4$ groups form sheets parallel to the $ac$ plane and stacked along the $b$-axis. Below the transition  at 373~K a spontaneous polarization arises in these sheets, with the neighboring sheets polarized in the opposite directions \cite{semmingsen:95,semmingsen:77,hollander:77}. 
At the transition the crystal symmetry changes from centrosymmetric tetragonal, $I4/m$, to monoclinic, $P2_1/m$.

Protons on the hydrogen bonds in squaric acid move in double-well potentials, so each of the protons can occupy a site closer to the   C$_4$O$_4$ group or the other site on the bond, closer to the neighboring C$_4$O$_4$ group. The C$_4$O$_4$ groups possess C$_{1h}$ symmetry both below above the transition \cite{moritomo:91}; the distortion from the square shape is caused by formation of a double $\pi$-bond between those two adjacent carbon atoms, closer to which two protons on the hydrogen bonds sit. The sublattice polarization is mostly electronic, while the direct contribution of ions and displaced protons is significantly smaller \cite{horiuchi:18}.

External electric fields induce many interesting phenomena in antiferroelectrics. Probably best known of them is switching of a negative sublattice polarization by the bias field and the resulting transition from antiferroelectric (AFE) to a ferroelectric (FE) phase, manifesting itself in the classical double $P-E$ hysteresis loops. Typical high dielectric permittivity and breakdown field of antiferroelectrics and small remnant polarization of these double hysteresis loops make these systems promising candidates for use in the high-energy-storage capacitors.

In conventional ferroelectrics the external electric field applied along the axis of spontaneous polarization is a field conjugate to the order parameter. The $T-E$ phase diagram in the case of the first order phase transition at zero fied is simple. The transition temperature increases with the field, and the transition line ends at the critical point \cite{stasyuk:01,kutnjak:07,novak:13,ma:16}. At the fields above the critical one, the rounded maxima of the dielectric permittivity are observed along the so-called Widom line \cite{kutnjak:07,novak:13,ma:16}. In the case of antiferroelectrics, there is no physically realizable electric field conjugate to the sublattice polarization, and the phase diagram is much more complicated.

Okada et al \cite{okada:74,suzuki:83,suzuki:83:2} and recently Tol\`{e}dano \cite{toledano:16} explored the $T-E$ phase diagram of a \textit{uniaxial} antiferroelectric using the phenomenological approach.  In particular, it has been found  \cite{okada:74,suzuki:83,suzuki:83:2} that apart from the  field-induced FE phase, there exists an additional semipolar phase at elevated fields, where an incomplete compensation of the sublattice polarizations occurs.

In contrast to the uniaxial antiferroelectrics, where the sublattice polarizations can be either parallel or antiparallel to the predetermined direction, the pseudo-tetragonal symmetry of squaric acid crystal lattice and, foremost, of its 2D hydrogen bond networks, permits also 90$^\circ$ rotations of the sublattice polarizations  by the applied electric field.  Hysteresis loop measurements and \textit{ab initio} calculations by Horiuchi et al \cite{horiuchi:18} have given evidence for such a rotation. Microscopically it would invoke a relocation of one of the two protons in each molecule to the other site along the same hydrogen bond and a simultaneous switching of the $\pi$ bond to those carbon atoms, close to which the protons would sit after the relocation. Further \textit{ab initio} calculations \cite{ishibashi:18} have shown that the field induced sublattice polarization rotation by 90$^\circ$ is possible at different orientations of the field within the $ac$ plane. It is also predicted \cite{horiuchi:18} that higher fields can lead to a relocation of the second proton along the hydrogen bond, resulting in a 180$^\circ$ rotation of the negative sublattice polarization as compared to the initial AFE phase and inducing the collinear ferroelectric phase, analogous to that in uniaxial antiferroelectrics.

Theoretical model description of the antiferroelectric transition in squaric acid is usually based on a certain version of the proton ordering model, invoking four-particle correlations between protons within the planes \cite{matsushita:80,matsushita:81,matsushita:82,chaudhuri:90}.
 The four-particle Hamiltonians are basically identical to those for NH$_4$H$_2$PO$_4$ crystals, antiferroelectrics of the KH$_2$PO$_4$ family. 
Recently we have developed a deformable modification of the four-particle model for squaric acid  \cite{moina:20} that describes the effects associated with the diagonal lattice strains and changes in the hydrogen bond geometry: the thermal expansion, influence of external hydrostatic pressure, and  dependence of the interaction constants on the H-site distance $\delta$.

To our best knowldedge, no theoretical study of the squaric acid behavior under external field influence at non-zero temperatures has been carried out yet. To that end in the present paper we shall modify our recently proposed model \cite{moina:20} to take into account the changes in the system symmetry caused by the electric fields applied within the plane of the hydrogen bonds. The system Hamiltonian, thermodynamic potential, and equations for the order parameters, polarization and lattice strains are obtained in section 2. Results of numerical calculations are presented in Section 3, where  the temperature-electric field phase diagram is constructed.
The Appendix contains a thorough coverage of the variety of the temperature dependences of the order parameters, strains, and static dielectric permittivity at different values of the field.

\section{The Model}
\begin{figure}[htb]
	\centerline{\includegraphics[width=\columnwidth]{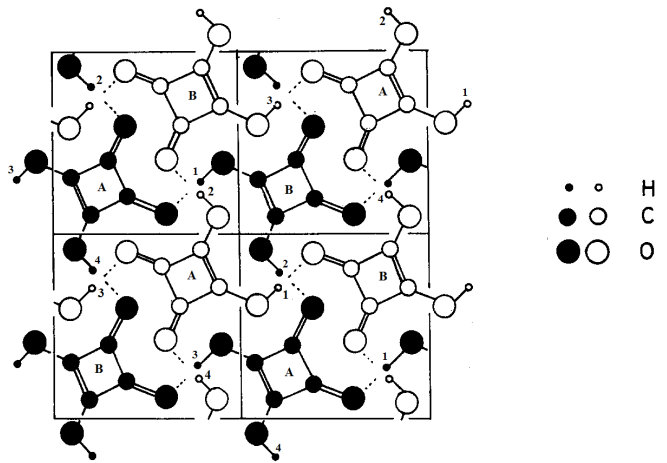}}
	\caption{Crystal structure of squaric acid as viewed along the $b$ axis. Figure is taken from \cite{semmingsen:74,moina:20}. Two adjacent layers are shown, with black and open circles each. The A and B type C$_4$O$_4$ groups are  indicated (see text for explanation), and the hydrogen bonds are numbered.} \label{sqa-structure}
\end{figure}

There are two formula units in the low-temperature phase unit cell of squaric acid. In our model the unit cell consists of two C$_4$O$_4$ groups and four hydrogen atoms ($f=1,2,3,4$, see fig.~\ref{sqa-structure}) attached to one of them (the A type group). All hydrogens around the B type groups are considered to belong to the A type groups, with which the B groups are hydrogen bonded. Note that the two C$_4$O$_4$ groups of each unit cell belong to
different neighboring layers. The center of each hydrogen bond lies exactly above   the center of the hydrogen bond in the layer below it (as seen along the $b$ axis).

As usually in the proton ordering models, we consider interactions between protons, leading to ordering in their system. Motion of protons in double-well potentials is described by pseudospins, whose two eigenvalues $\sigma=\pm 1$ are assigned to two equilibrium positions of the proton. It is assumed that the crystal is placed into electric fields $E_1$ and $E_3$, confined to the $ac$ plane.

The system Hamiltonian \cite{moina:20}
\begin{equation}\label{sqa-Ham}
H=U_{\textrm{seed}}+H_{\textrm{long}}^{\textrm{intra}}+H_{\textrm {long}}^{\textrm{inter}}+H_{\textrm{short}}
\end{equation}
includes ferroelectric intralayer long-range interactions $H_{\textrm{long}}^{\textrm {intra}}$, ensuring ferroelectric 
ordering within each separate layer, antiferroelectric interlayer $H_{\textrm{ long}}^{\textrm{inter}}$ responsible for alternation of polarizations in the stacked layers, the short-range configurational interactions between protons $H_{\textrm{short}}$, which includes also the interactions with external electric fields, and the so-called ``seed'' energy \cite{moina:20}
\begin{equation}\label{key2}
U_{\textrm{seed}}=vN\left[\frac 12\sum_{ij=1}^3c_{ij}^{(0)}\eps_i\eps_j-\sum_{ij=1}^3c_{ij}^{(0)}\alpha_i^{(0)}(T-T_i^0)\eps_j\right],
\end{equation}
containing elastic and thermal expansion contributions associated with uniform diagonal lattice strains $\eps_i$; $c_{ij}^{(0)}$ and $\alpha_i^{(0)}$ are the ``seed'' elastic constants and thermal expansion coefficients; $T_i^0$ determine the reference point of the thermal expansion of the crystal, which can be chosen arbitrarily; $v$ is the unit cell volume, and $N$ is the number of the unit cells in the crystal. 
Contributions of the monoclinic strain $\eps_5$ are ignored.

\begin{figure}[htb]
	\centerline{\includegraphics[angle=90,height=0.19\textwidth]{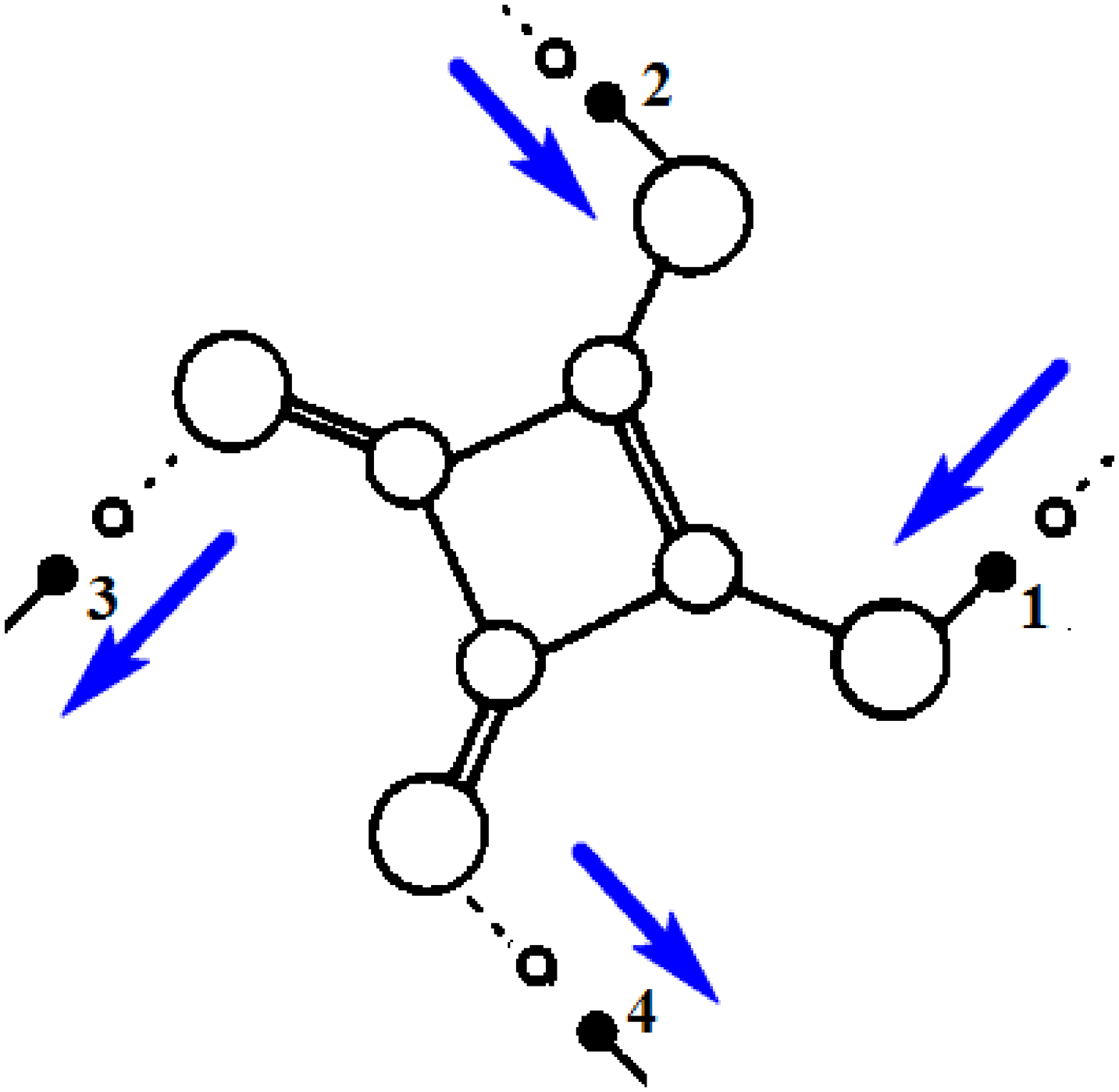}~~~~~~~~~~~~~~
	\includegraphics[height=0.18\textwidth]{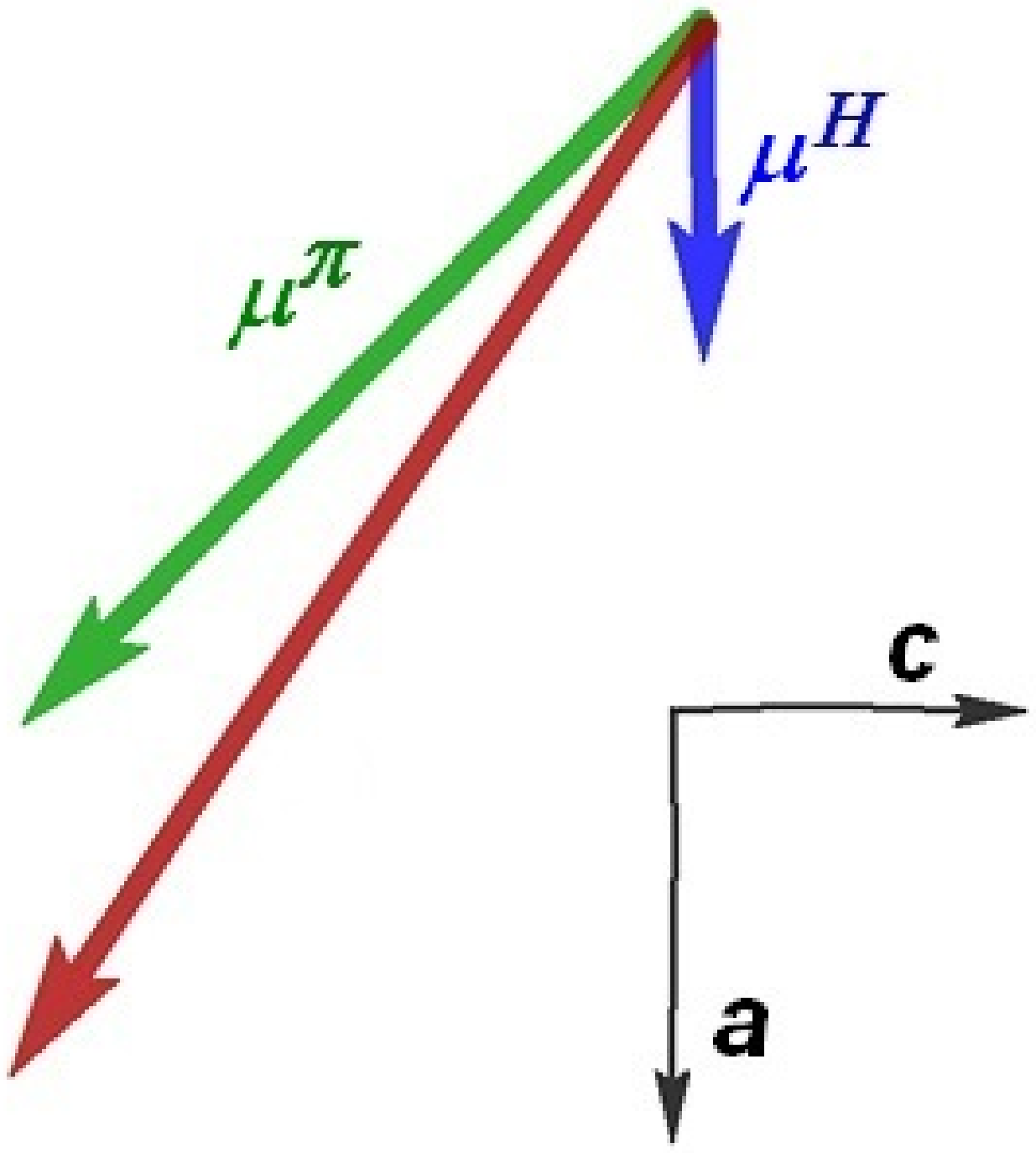}}
	\caption{Dipole moments of the lateral proton configuration (configuration 1 of Table~\ref{configurations_table}) around an A type C$_4$O$_4$ group, as deduced from the results of \cite{horiuchi:18}.  ${\bm{\mu}}^H_1=(2\mu^H,0,0)$,
		$\bm{\mu}^\pi_1=(2\mu^\pi_\parallel,0,-2\mu^\pi_\perp)$.  Directions of the dipole moments associated with protons and with electrons are shown by blue and green arrows, respectively; the vector lengths are nominal. 
} \label{configurations_fig1}
\end{figure}

The short-range Hamiltonian $H_{\textrm{short}}$ describes the four-particle configurational correlations between protons sitting around each C$_4$O$_4$ group. 
It is assumed that the energy of four lateral configurations $\eps_a$, where two protons are in positions close to the adjacent oxygens of the C$_4$O$_4$ group, whereas two other protons are closer to the neighboring C$_4$O$_4$ groups (see fig.~\ref{configurations_fig1} and Table~\ref{configurations_table}, configurations 1-4), is the lowest of all. This level is thus four time degenerate  in absence of external electric field. The next level is two diagonal configurations with the energy $\eps_s$, where the protons are close to the opposite oxygens of the C$_4$O$_4$ group (configurations 5-6 in Table~\ref{configurations_table}). Then there are eight single-ionized configurations with three or only one close protons, having the energy $\eps_1$ (configurations 7-14), and two double-ionized configurations with four or no proton at all close to the given C$_4$O$_4$ group ($\eps_0$, configurations 15-16). We accept that $\eps_a<\eps_s\ll\eps_1\ll\eps_0$.

The external electric fields $E_1$ and $E_3$ remove the degeneracy of the lowest level of four lateral proton configurations $\eps_a$, as well as of the second excited level of single-ionized configurations $\eps_1$.
The energies of the diagonal (5-6) and double-ionized (15-16) proton configurations, having no dipole moment in the  $ac$ plane, remain unchanged. 

Our modeling of the electric field influence on the squaric acid
is heavily based on the results of the Berry phase calculations \cite{horiuchi:18} by Horiuchi et al, who have shown that the sublattice polarization in this crystal is formed predominantly by the electronic contributions of switchable $\pi$-bond dipoles, rather than directly by displacements  of protons along the hydrogen bonds. Positions of the double bonds, however, are determined by the proton configurations around the given C$_4$O$_4$ group: in the ground state lateral configurations the bond is formed between the two neighboring carbons, near which protons sit on the hydrogen bonds (see fig.~\ref{configurations_fig1}), and also between the carbons and adjacent to them oxygens, next to which there is no proton (meaning that the protons on these bonds sit in the minima close to the neighboring C$_4$O$_4$ groups). 

Directions of the electronic and total sublattice polarizations are basically determined by populations of the four ground state lateral proton configurations in the sublattice. For the configuration 1 these directions are  depicted in fig.~\ref{configurations_fig1}. The proton contribution vector goes along the [100] axis, while the electronic contribution is at an angle to it.  The three other  lateral  configurations and their dipole moments can be obtained from the scheme of fig.~\ref{configurations_fig1} by rotation by a multiple of 90$^\circ$. It is taken that in absence of electric field the protons are mostly in the configuration 1  in the positively polarized sublattice and in the configuration 3 in the negatively polarized sublattice. Then, the observed \cite{horiuchi:18}  90$^\circ$ switching of polarization means that the protons in negatively polarized sublattice move from the configuration 3 to configuration 2, and the total 180$^\circ$ rotation of polarization means switching to the configuration 1.

\begin{figure}[htb]
	\centerline{\includegraphics[angle=90,height=0.2\textwidth]{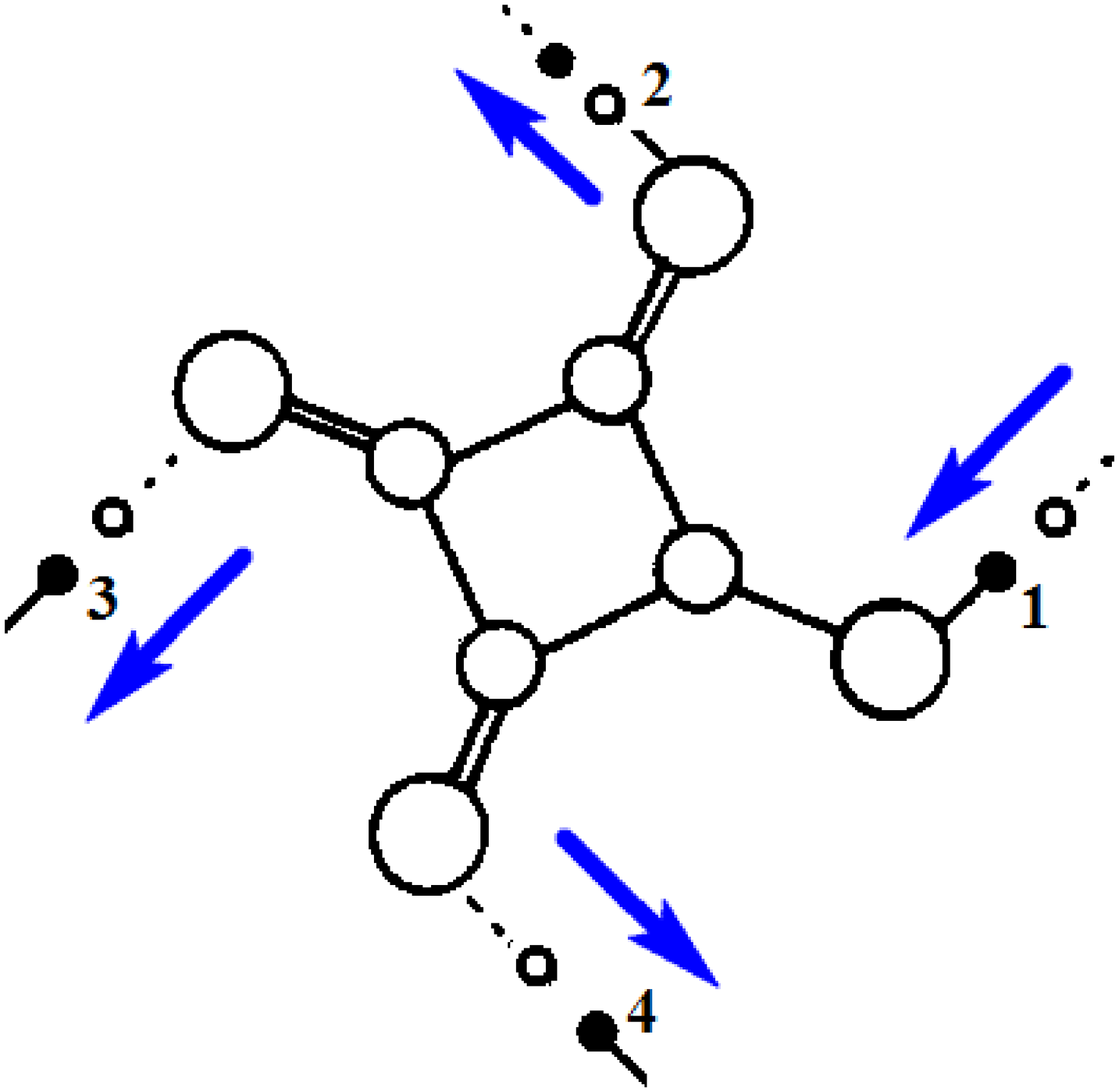}~~~~
		\includegraphics[angle=90,height=0.15\textwidth]{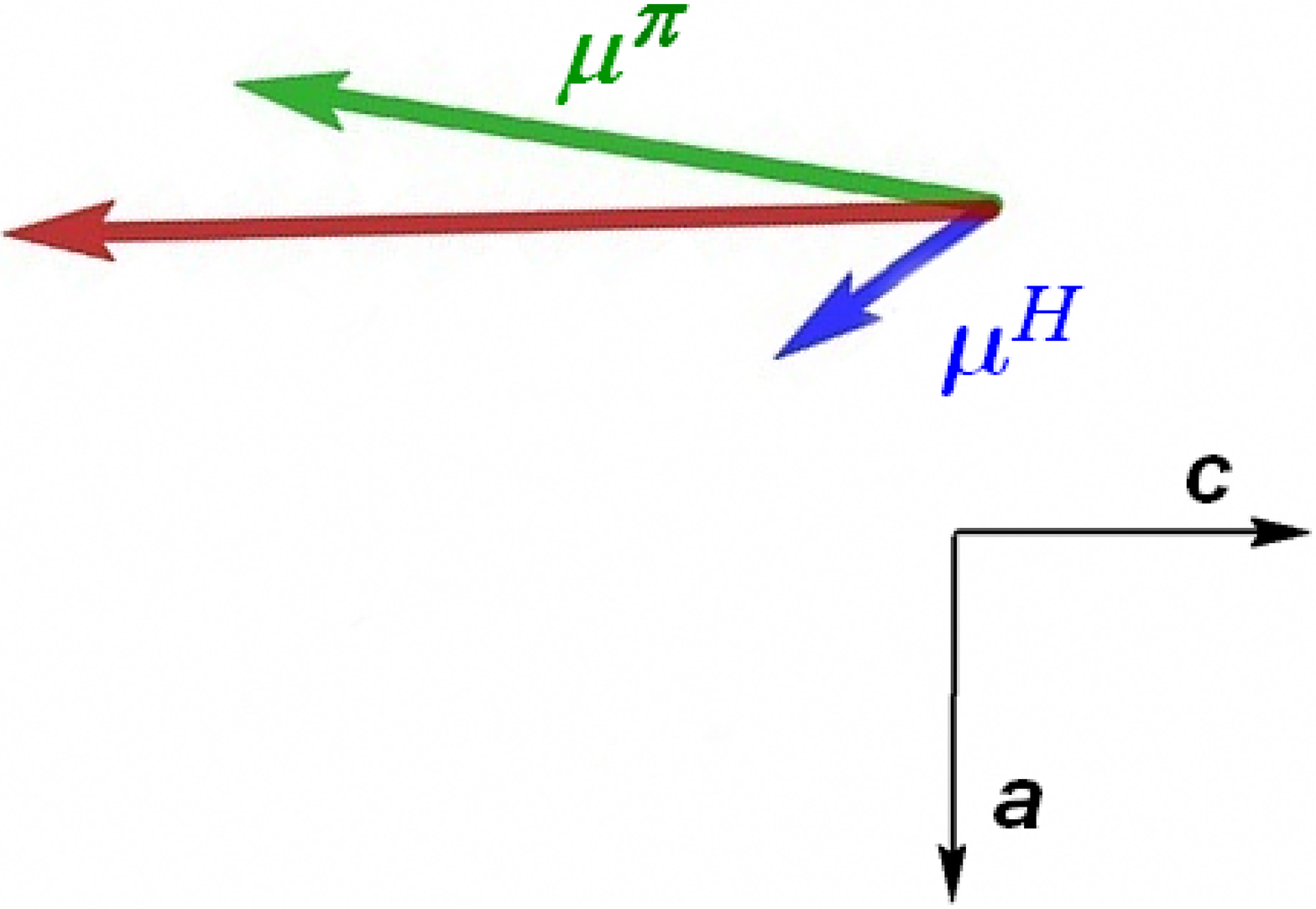}}
	\caption{Same as in fig.~\ref{configurations_fig1} for the single-ionized  proton configuration with one proton (configuration 7 of Table~\ref{configurations_table}). ${\bm{\mu}}^H_7=(\mu^H,0,-\mu^H)$,
		$\bm{\mu}^\pi_7=(\mu^\pi_\parallel-\mu^\pi_\perp,0,-\mu^\pi_\parallel-\mu^\pi_\perp)$.  } \label{configurations_fig2}
\end{figure}

No polarization calculations for the single-ionized configurations, like the Berry phase calculations for the lateral configurations \cite{horiuchi:18} has been performed yet. Hence, while the direction of the protonic polarization can be easily determined, the direction of the electronic polarization for them is largely a guesswork.  As one can see in fig.~\ref{configurations_fig1}, for the lateral configurations the electronic dipole moment is directed, roughly, from the two protons towards the two double bonds between carbons and oxygens, next to which there is no proton. For the symmetry reasons and in analogy to the case of the lateral configurations, we shall assume that in the single-ionized configuration (fig.~\ref{configurations_fig2}), the electronic dipole moment is directed from the single proton towards the double bond between the oxygen and carbon along the diagonal of the C$_4$O$_4$ group.
With the proton dipole moment directed at 45$^\circ$ to the crystallographic axes, we again obtain approximately the same angle between the the protonic and electronic polarization vectors, as in the case of the lateral configurations.

For the single-ionized configurations 11-14 with three protons, we use the same reasoning as above for the configurations 7-10 with a single proton. In particular, the dipole moments of the configuration 13 are the same as
those of the configuration 7. 

So, we assume that the angle between proton and electronic dipole moments is the same for all  lateral and single-ionized configurations, and \textit{ditto} for the ratio of the absolute values of the proton and electronic dipole moments.
For the lateral configurations the absolute value of the proton dipole moment is denoted as 2$\mu^H$; the electronic dipole moment vector has two components: 2$\mu^\pi_\parallel$ along the direction of the proton dipole moment and 2$\mu^\pi_\perp$ in the perpendicular direction. So,  if  we have ${\bm{\mu}}^H_1=(2\mu^H,0,0)$,
$\bm{\mu}^\pi_1=(2\mu^\pi_\parallel,0,-2\mu^\pi_\perp)$ for the configuration 1 and  ${\bm{\mu}}^H_7=(\mu^H,0,-\mu^H)$ for the
configuration 7, then from these assumptions it follows that 
$\bm{\mu}^\pi_7=(\mu^\pi_\parallel-\mu^\pi_\perp,0,-\mu^\pi_\parallel-\mu^\pi_\perp)$.
The directions and components of the dipole moments of the other lateral and single-ionized configurations and their dipole moments can be obtained by rotation of the configurations 1, 7, and 13. These directions and the configuration energies in presence of the electric fields $E_1$ and $E_3$
are summarized in Table~\ref{configurations_table}.

\renewcommand{\tabcolsep}{3.0pt}
\renewcommand{\arraystretch}{1.3}

\begin{table}[!h]
	\caption{Proton configurations and their energies in presence of the electric fields $E_1$ and $E_3$; $W_3=\mu^HE_1+\mu^\pi_{\parallel}E_1-\mu_\perp^\pi E_3$; 
	$W_1=-\mu^HE_3-\mu^\pi_{\parallel}E_3-\mu_\perp^\pi E_1$. Directions of the dipole moments associated with protons and with electrons are shown by blue and green arrows, respectively. }
	\label{configurations_table}
	\begin{center}
		\small
		\begin{tabular}{c|c|c|c|c}
	\hline
	$i$ & & $s_1s_2s_3s_4$ &&${\cal E}_i$ \\
	\hline
	1 & \includegraphics[height=1.0cm,width=1.0cm]{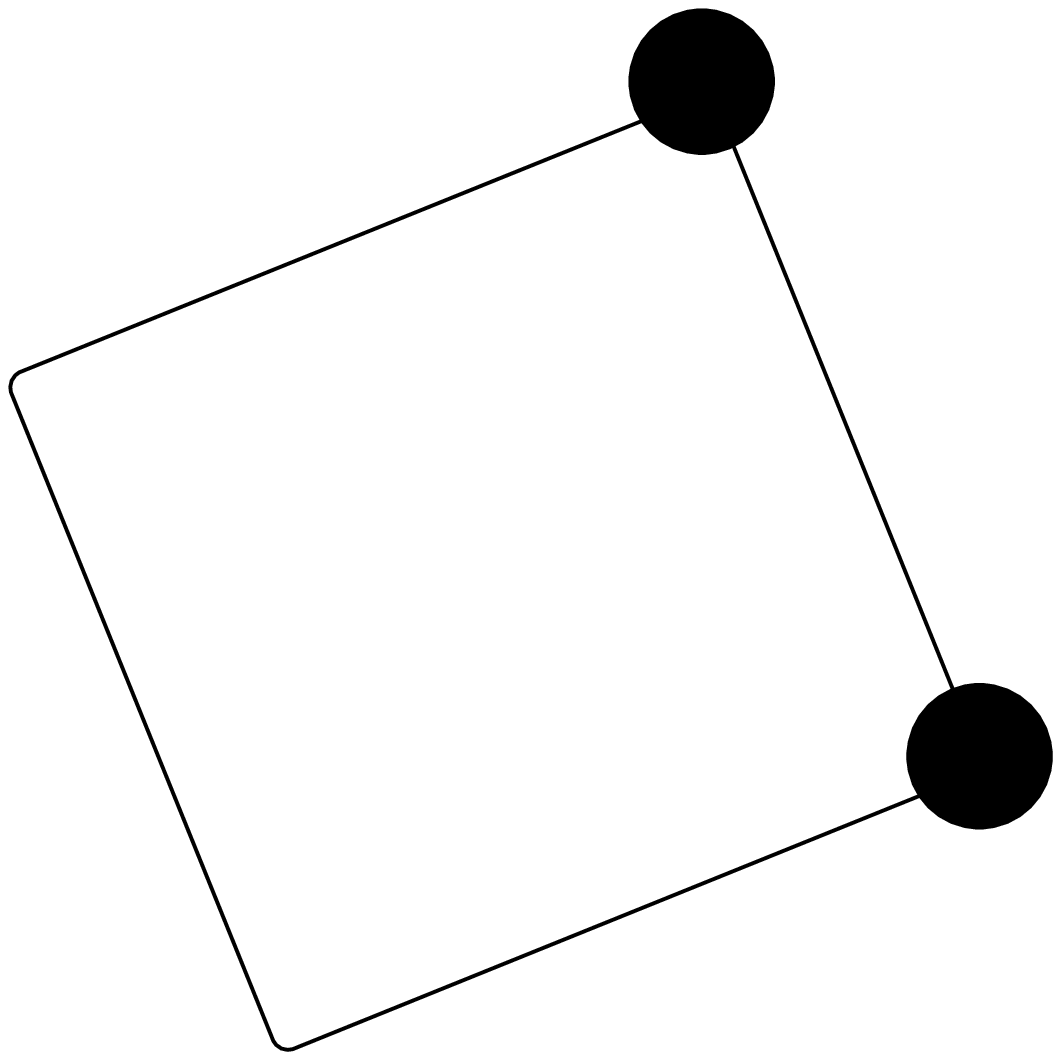} &   $++--$ &  \includegraphics[height=0.6cm,width=0.6cm]{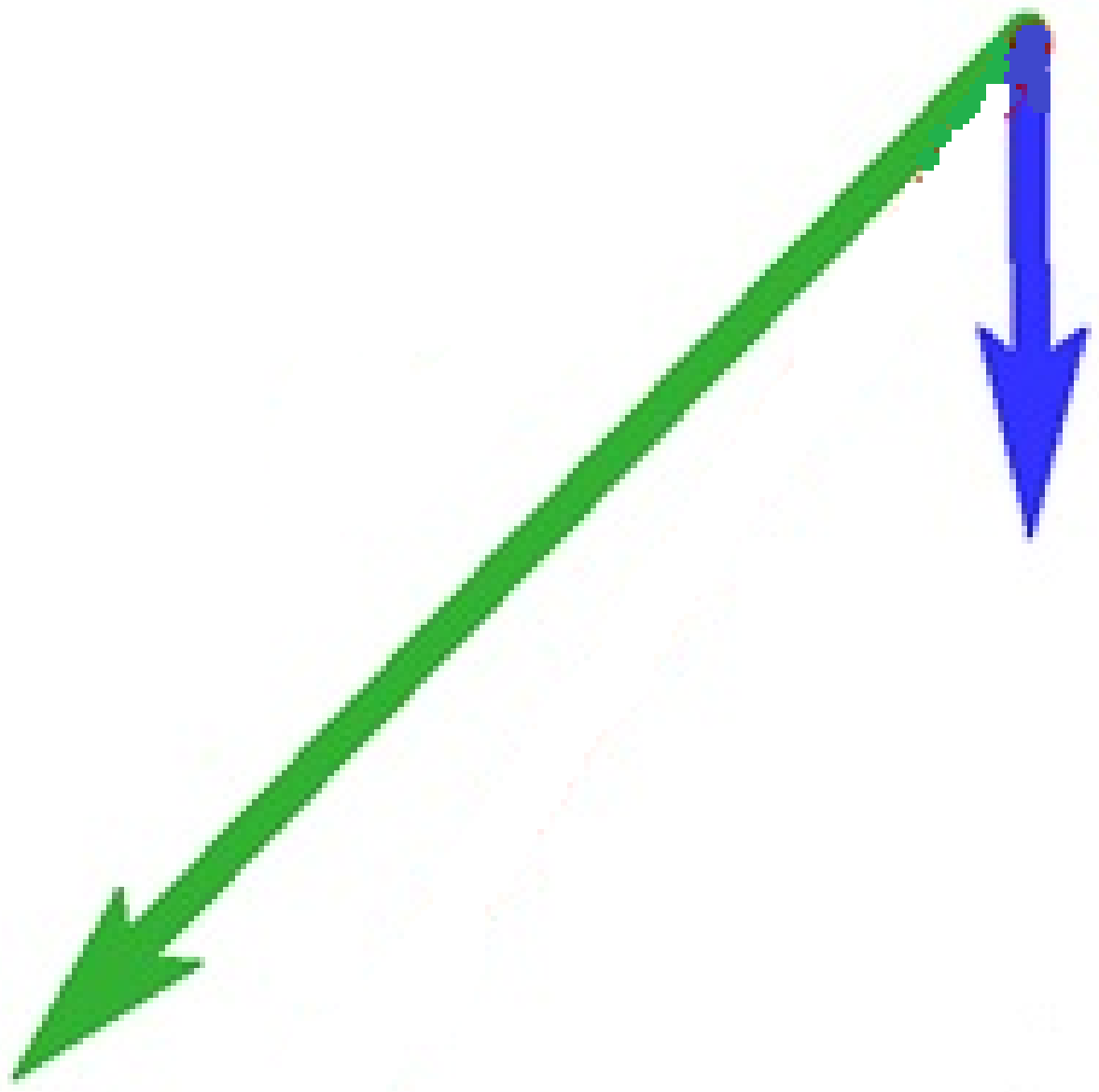} & $\eps_a - 2W_3$ \strut \\
	2 & \includegraphics[height=1.0cm,width=1.0cm]{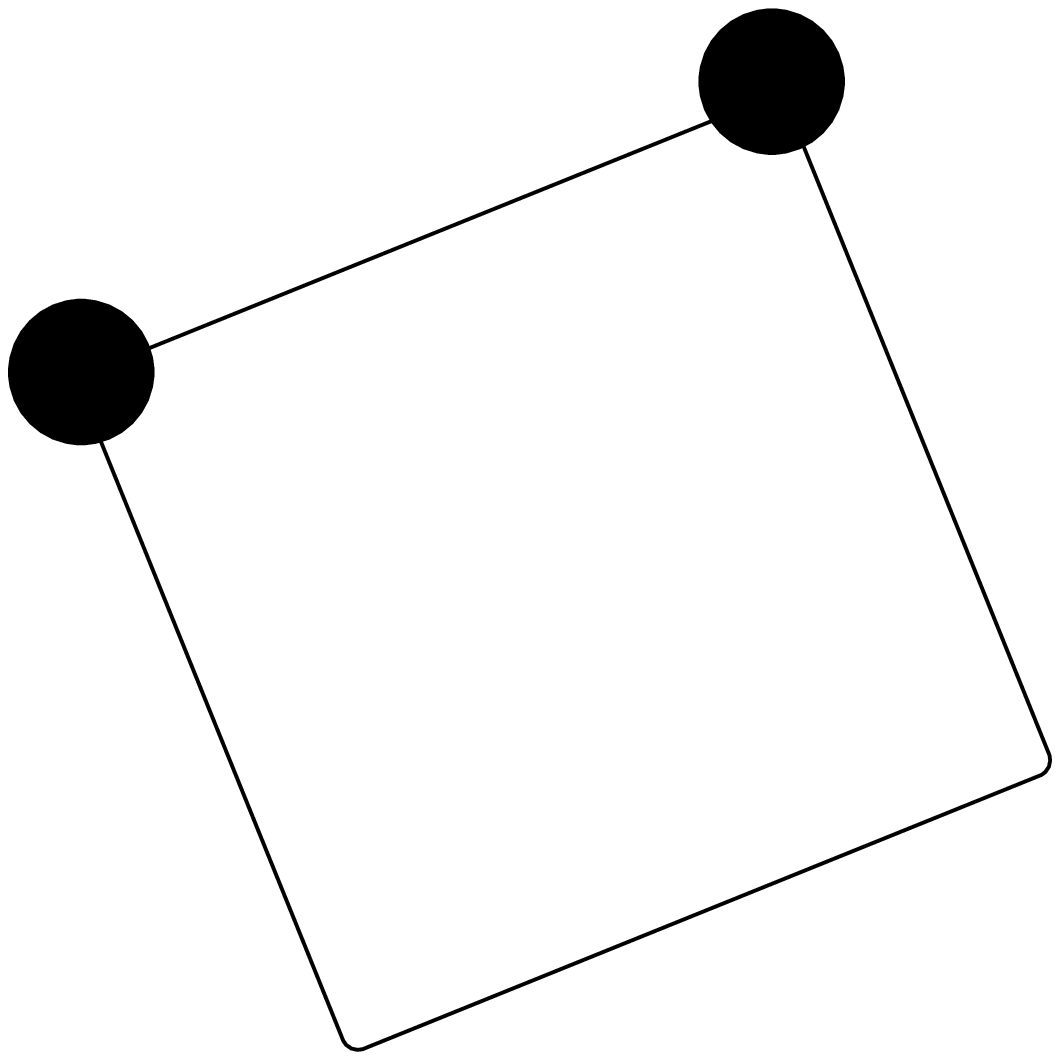} &   $-++-$ 
	&  \includegraphics[height=0.6cm,width=0.6cm,angle=90]{arrows1.eps} &
	$\eps_a +2W_1$  \\
	3 & \includegraphics[height=1.0cm,width=1.0cm]{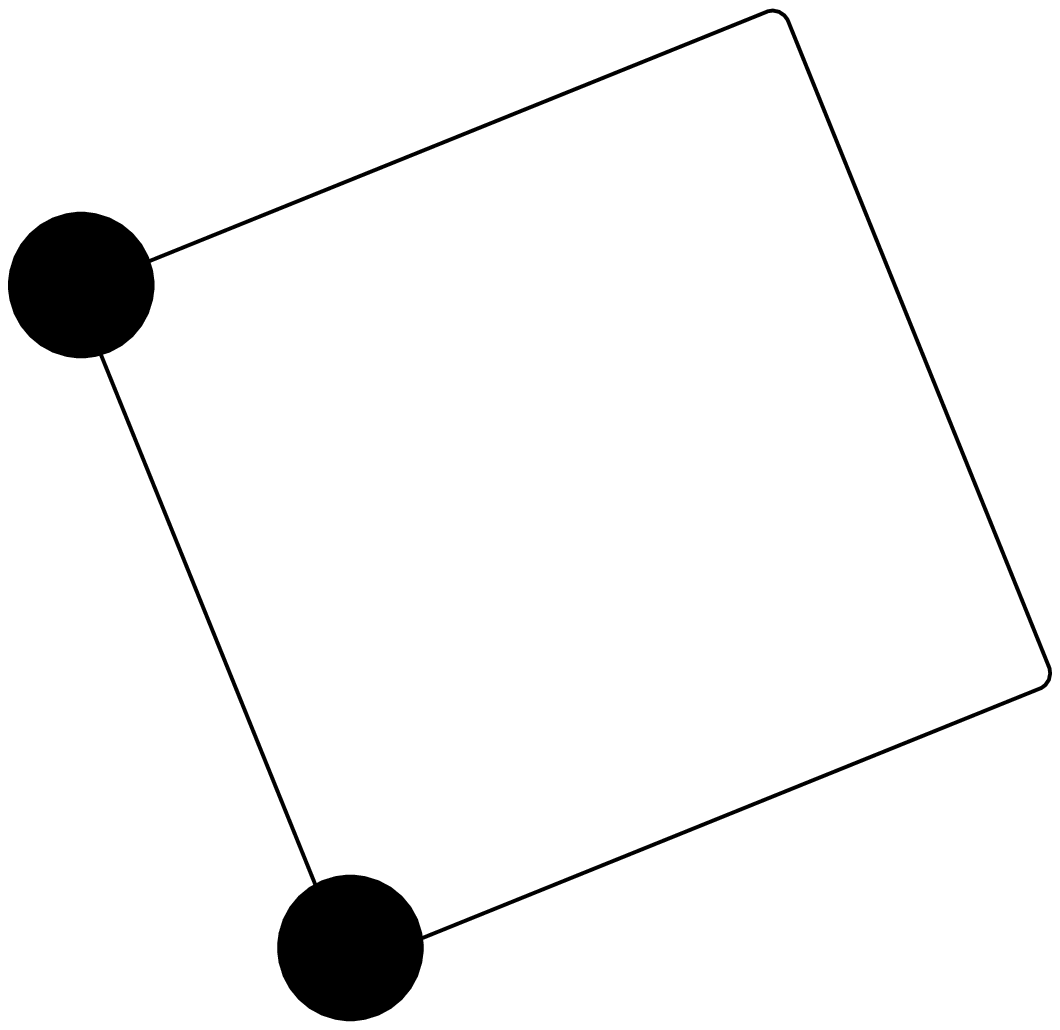} &   $--++$&  \includegraphics[height=0.7cm,width=0.7cm,angle=180,origin=c]{arrows1.eps} &
	$\eps_a + 2W_3$ \\
	4 & \includegraphics[height=1.0cm,width=1.0cm]{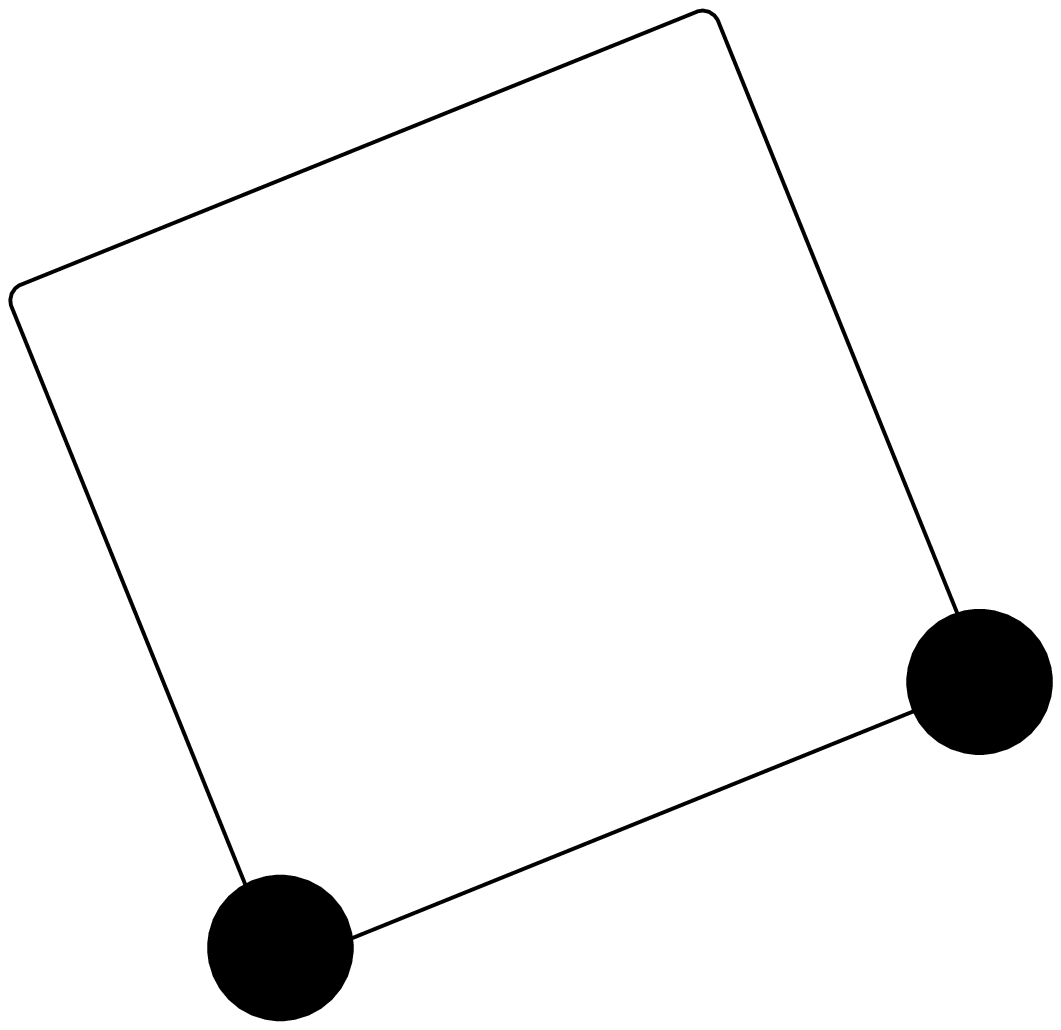} &  $+--+$ &
	\includegraphics[height=0.7cm,width=0.7cm,angle=270,origin=c]{arrows1.eps} &
		$\eps_a -2W_1$   \\
	\hline
	5 & \includegraphics[height=1.0cm,width=1.0cm]{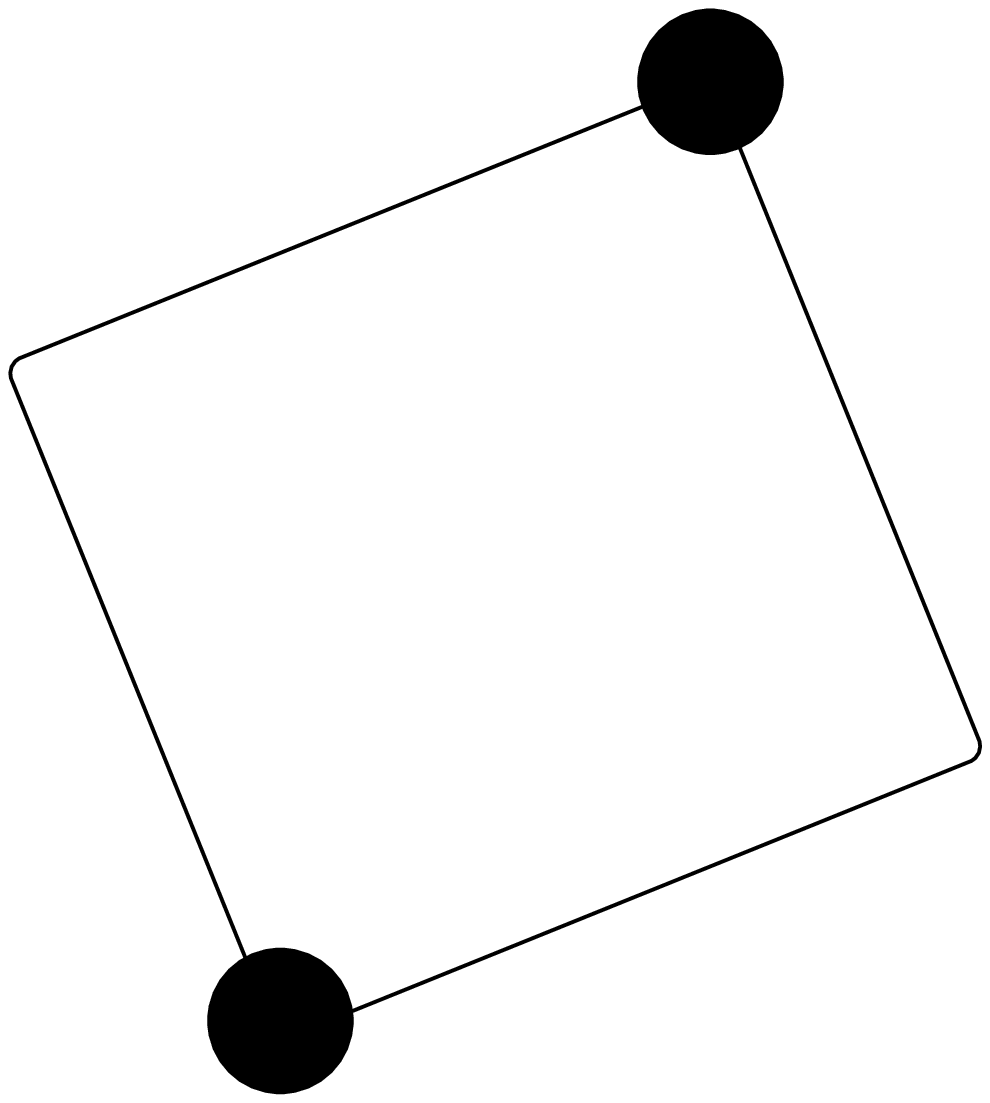} &  $-+-+$ & & $\eps_s$ \\
	6 & \includegraphics[height=1.0cm,width=1.0cm]{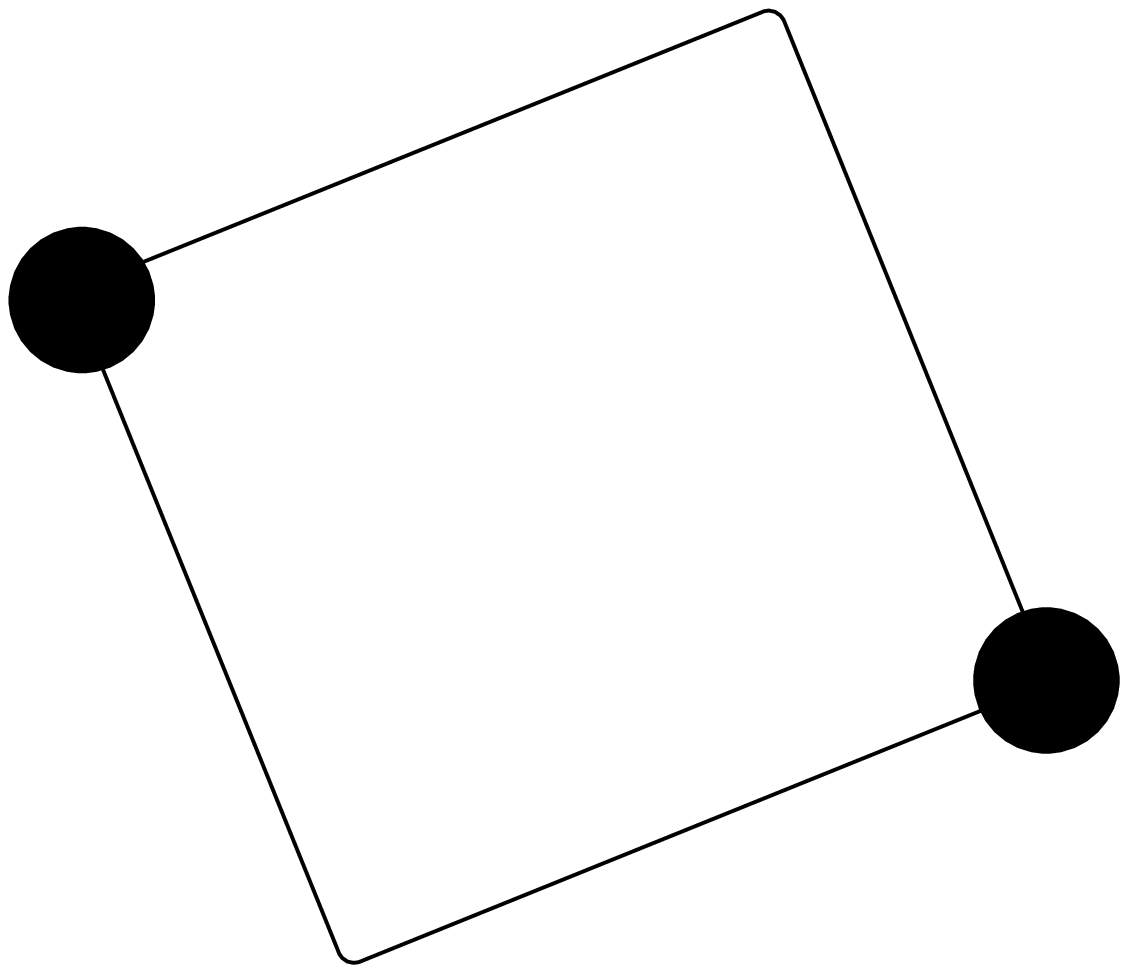} &  $+-+-$ & & $\eps_s$\\
	\hline
	7 & \includegraphics[height=1.0cm,width=1.0cm]{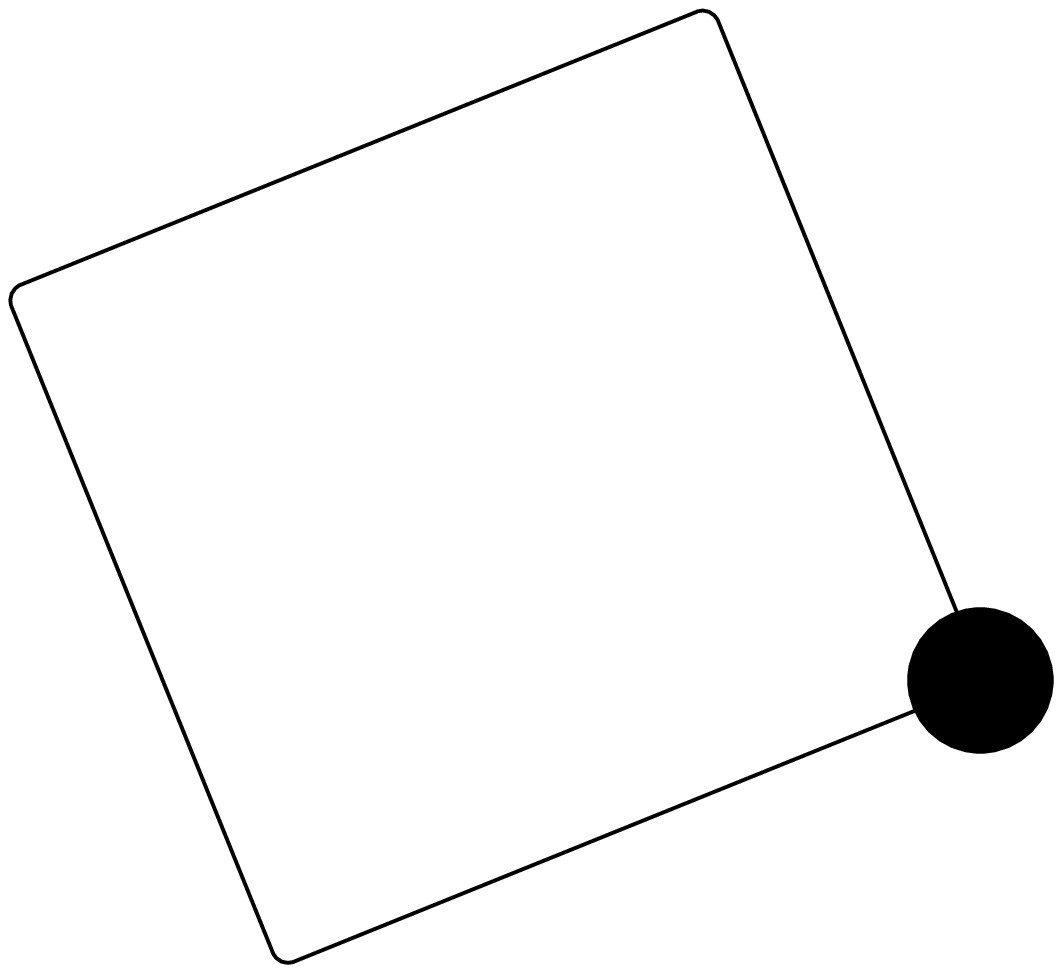}&   $+---$ &   \includegraphics[height=0.6cm,width=0.6cm,angle=-45,origin=c]{arrows1.eps} &$\eps_1-W_1-W_3$ \\
	8 & \includegraphics[height=1.0cm,width=1.0cm]{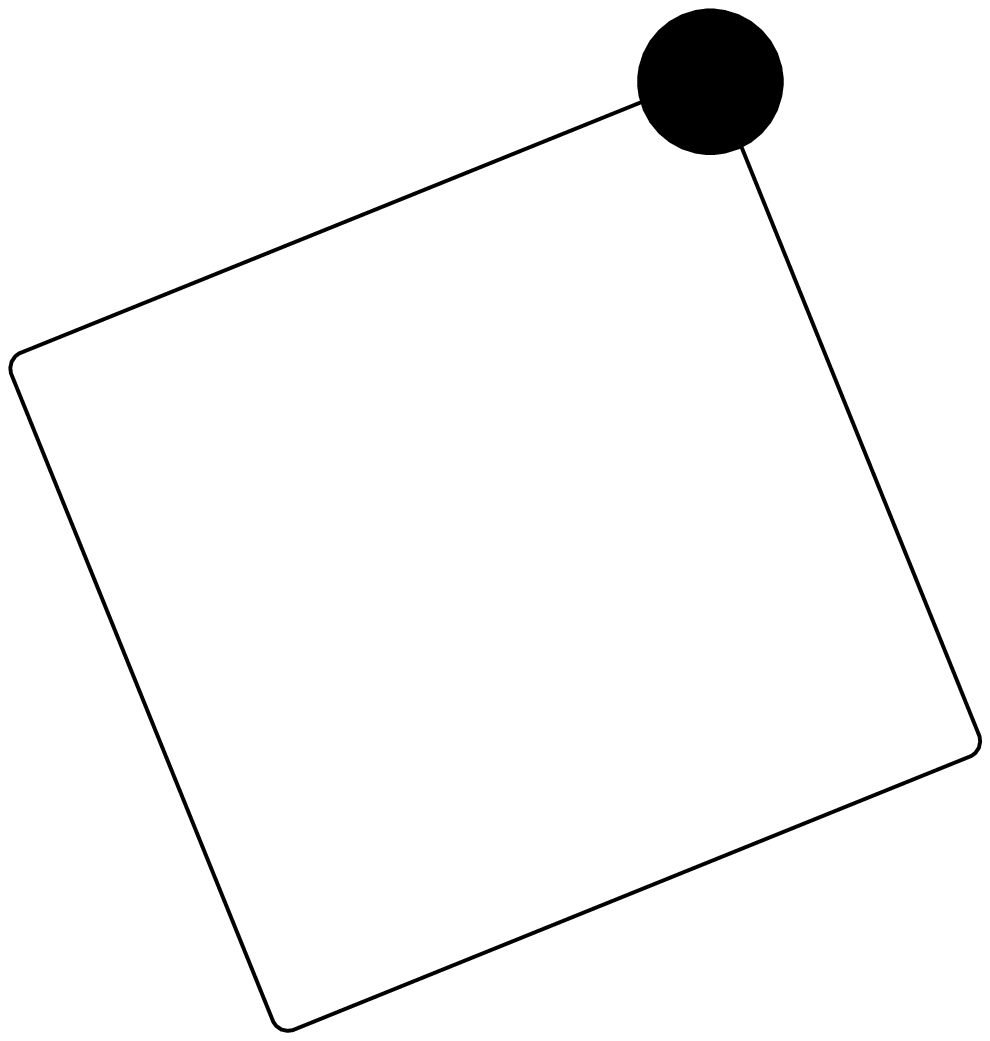}&  $-+--$ &  \includegraphics[angle=45,origin=c,height=0.7cm,width=0.7cm]{arrows1.eps} &$\eps_1+W_1-W_3$ \\
	9 & \includegraphics[height=1.0cm,width=1.0cm]{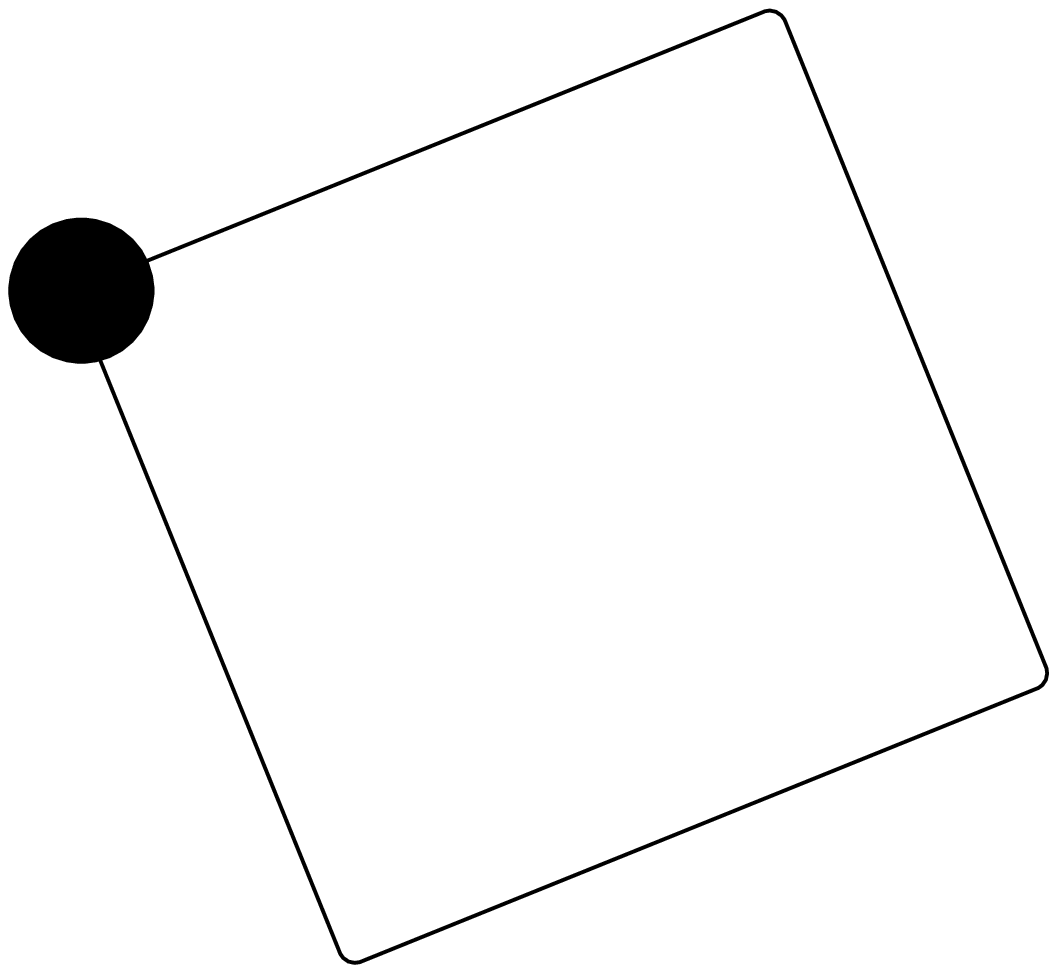} &  $--+-$ &
	  \includegraphics[angle=135,origin=c,height=0.7cm,width=0.7cm]{arrows1.eps} & $\eps_1+W_1+W_3$   \\
	10 & \includegraphics[height=1.0cm,width=1.0cm]{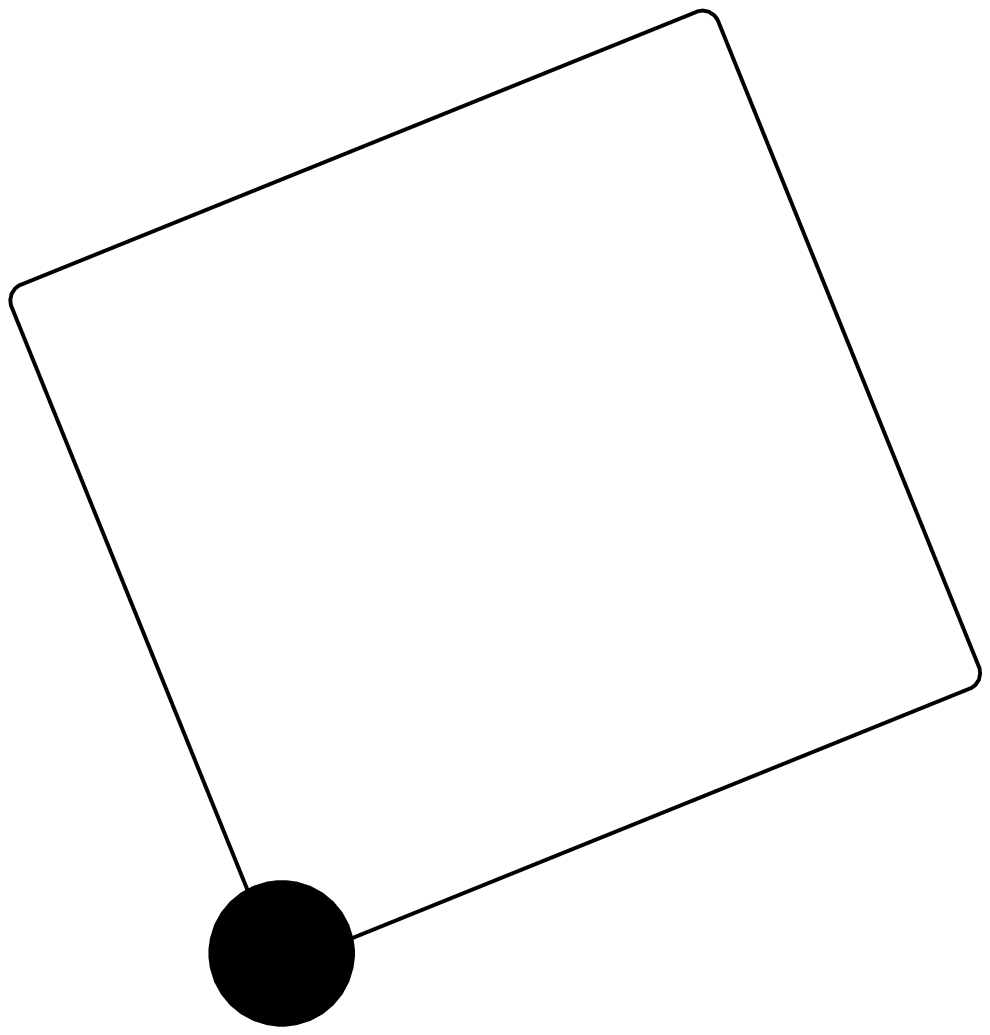} & $---+$ &    \includegraphics[angle=225,origin=c,height=0.7cm,width=0.7cm]{arrows1.eps}  & $\eps_1-W_1+W_3$\\
	11 & \includegraphics[height=1.0cm,width=1.0cm]{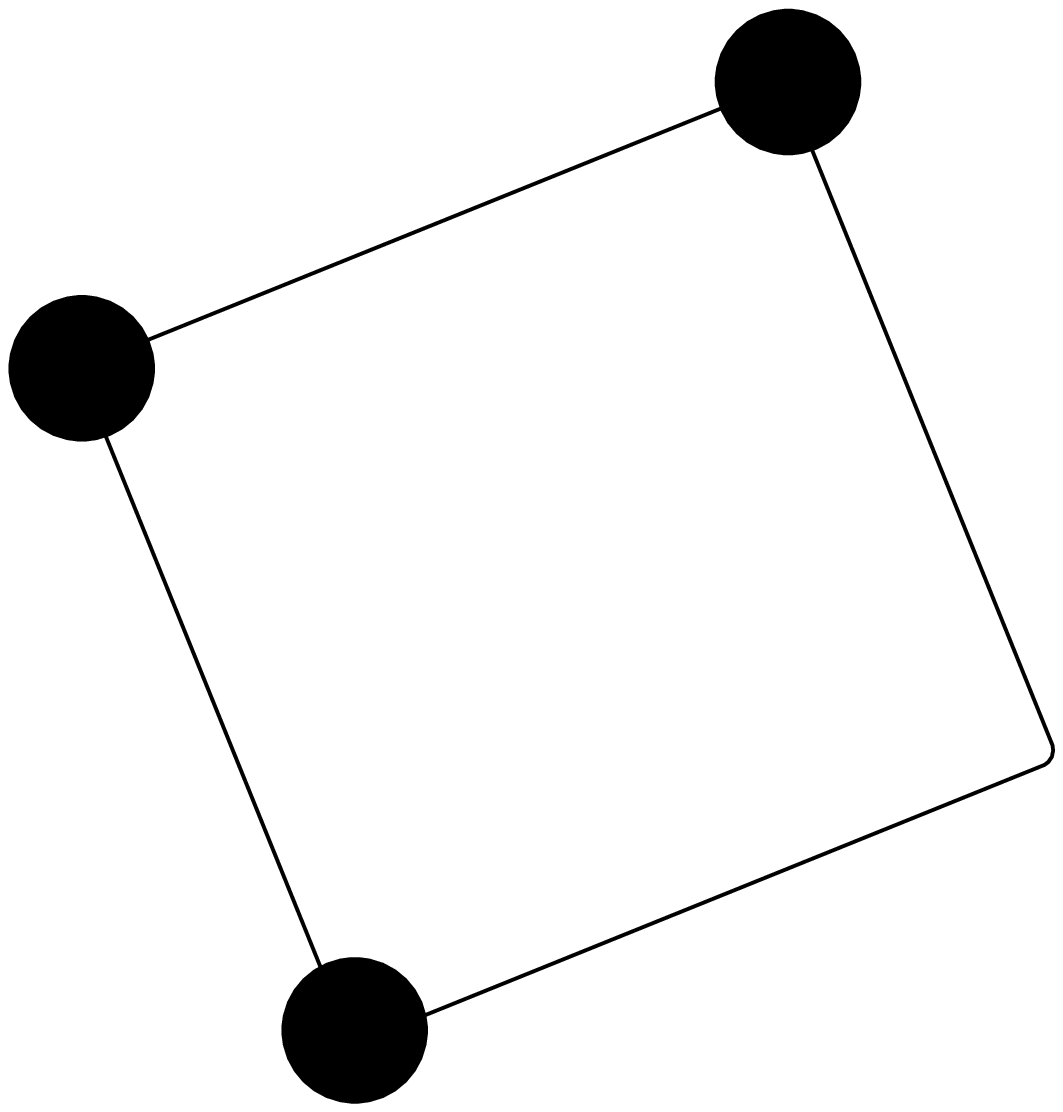} &   $-+++$ &    \includegraphics[angle=135,origin=c,height=0.7cm,width=0.7cm]{arrows1.eps} &  $\eps_1+W_1+W_3$ \\
	12 & \includegraphics[height=1.0cm,width=1.0cm]{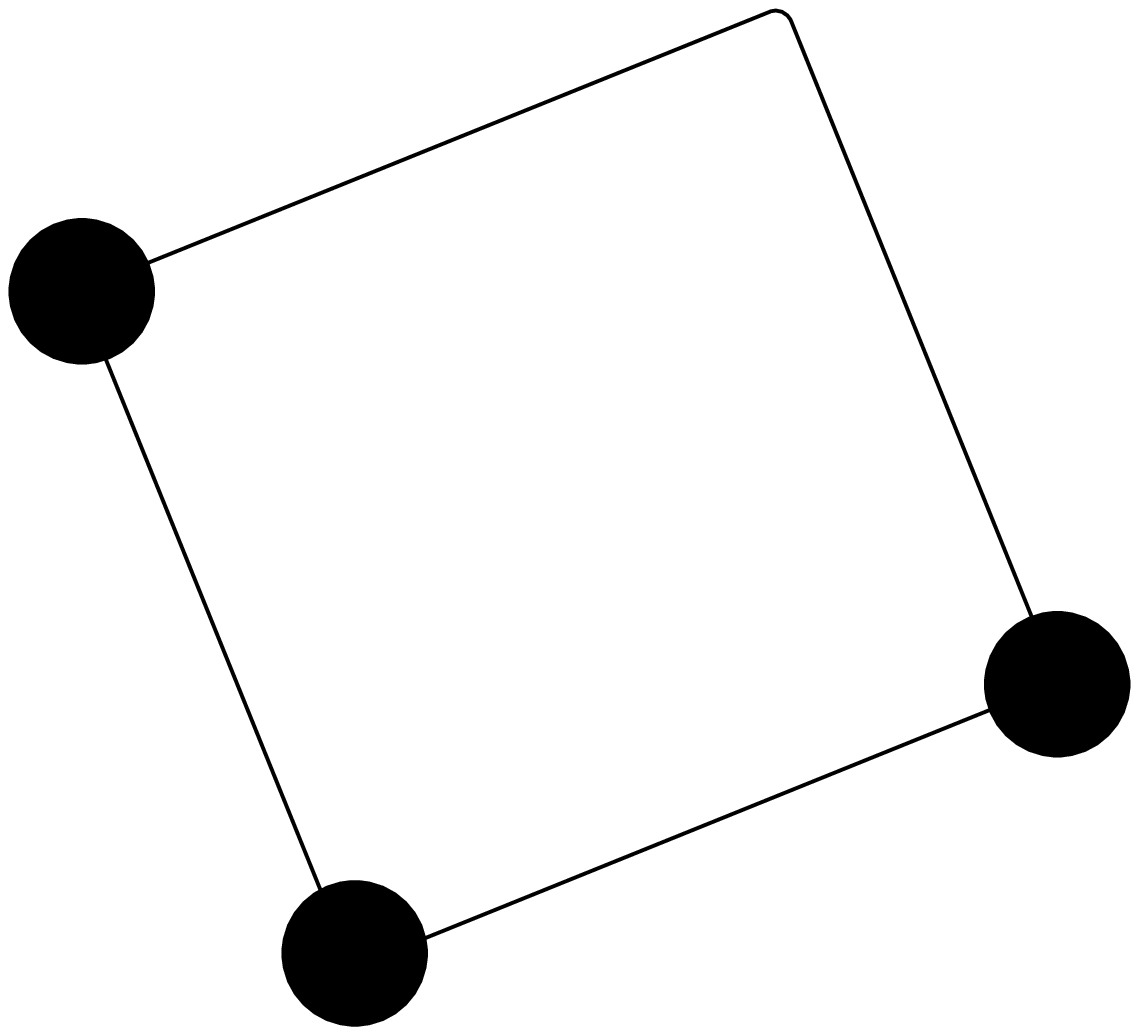} &   $+-++$&    \includegraphics[angle=225,origin=c,height=0.7cm,width=0.7cm]{arrows1.eps} & $\eps_1-W_1+W_3$ \\
	13 & \includegraphics[height=1.0cm,width=1.0cm]{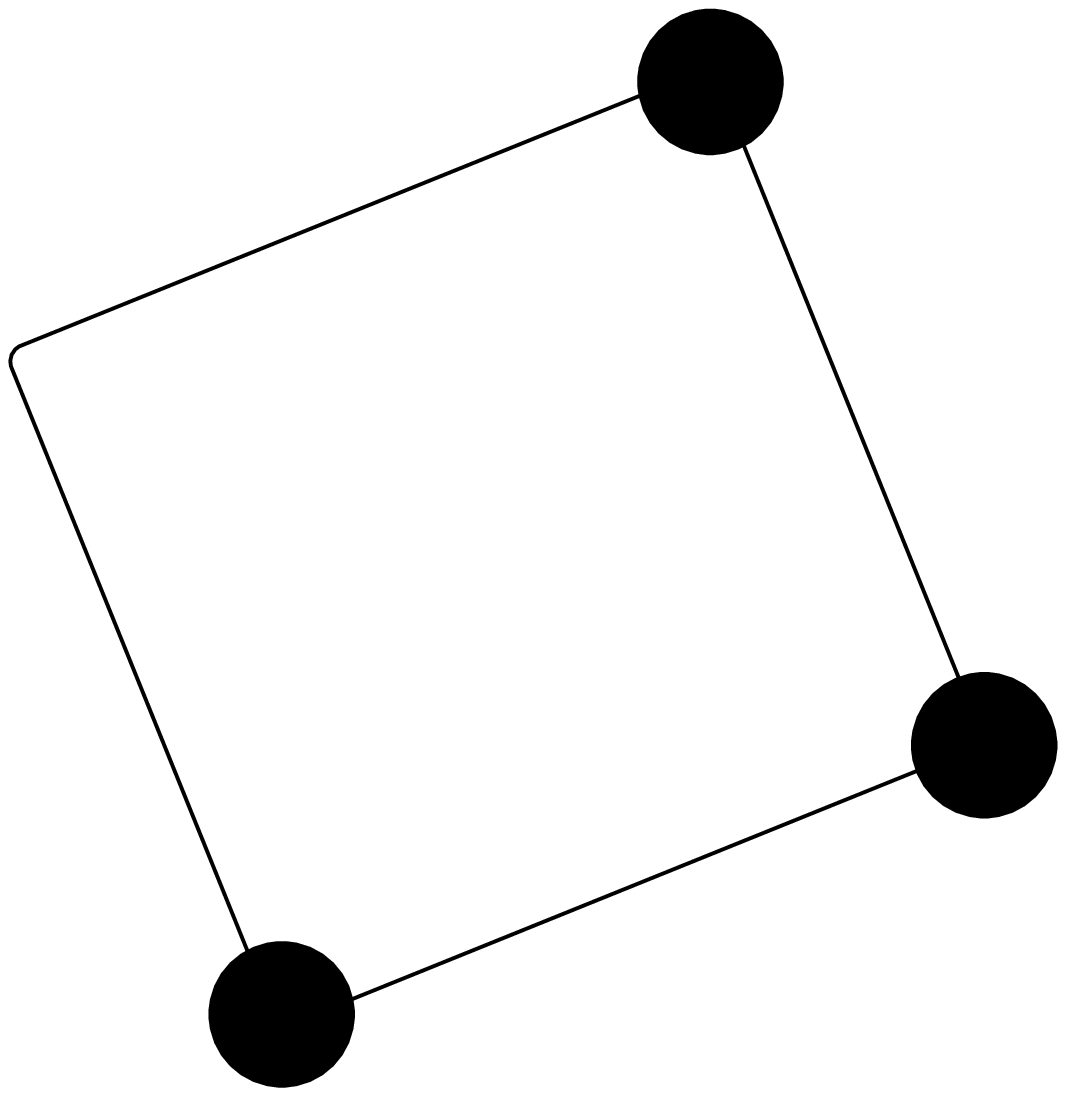} &   $++-+$ &    \includegraphics[angle=-45,origin=c,height=0.7cm,width=0.7cm]{arrows1.eps} & $\eps_1-W_1-W_3$\\
	14 & \includegraphics[height=1.0cm,width=1.0cm]{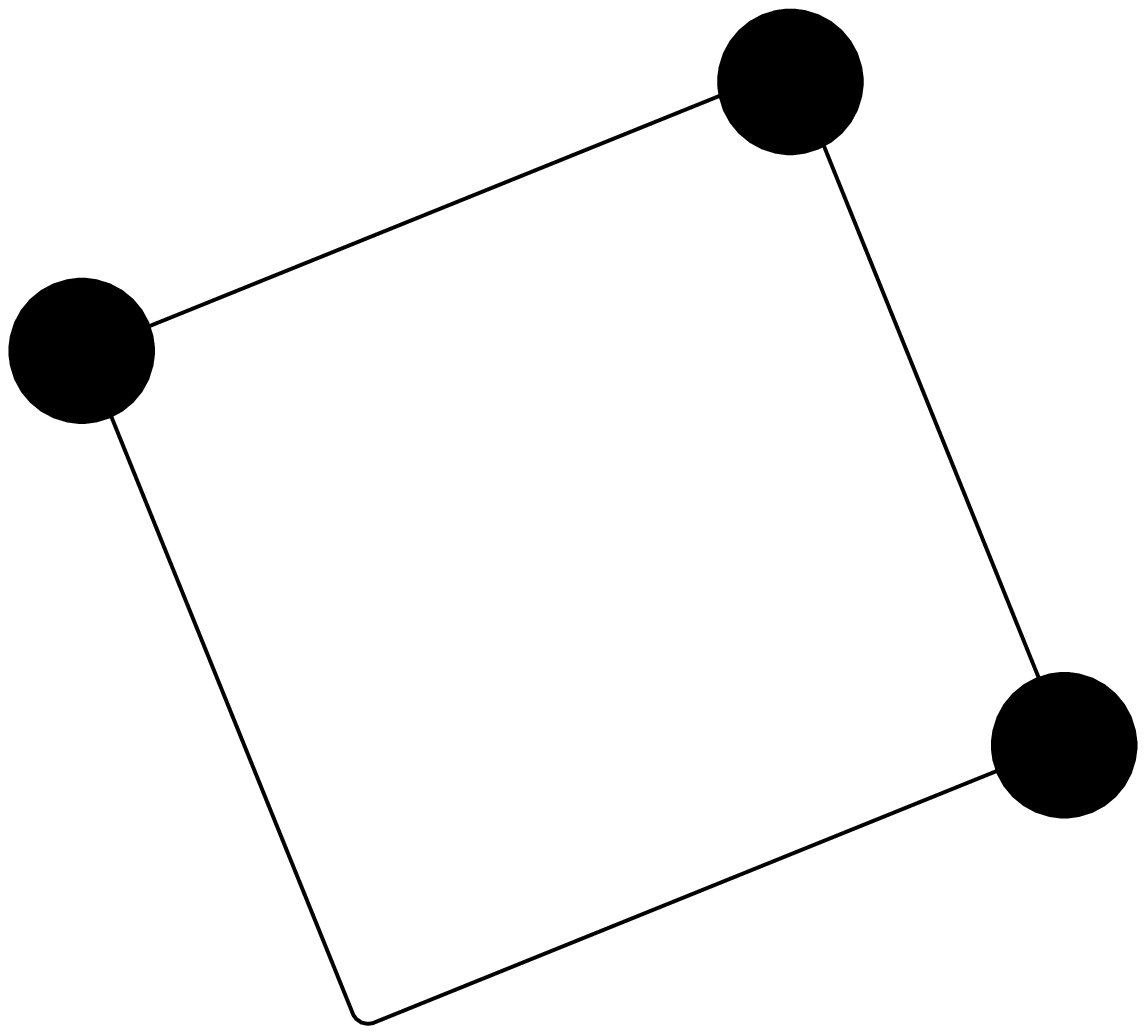} &   $+++-$ &    \includegraphics[angle=45,origin=c,height=0.6cm,width=0.6cm]{arrows1.eps} & $\eps_1+W_1-W_3$ \\
	\hline
	15 & \includegraphics[height=1.0cm,width=1.0cm]{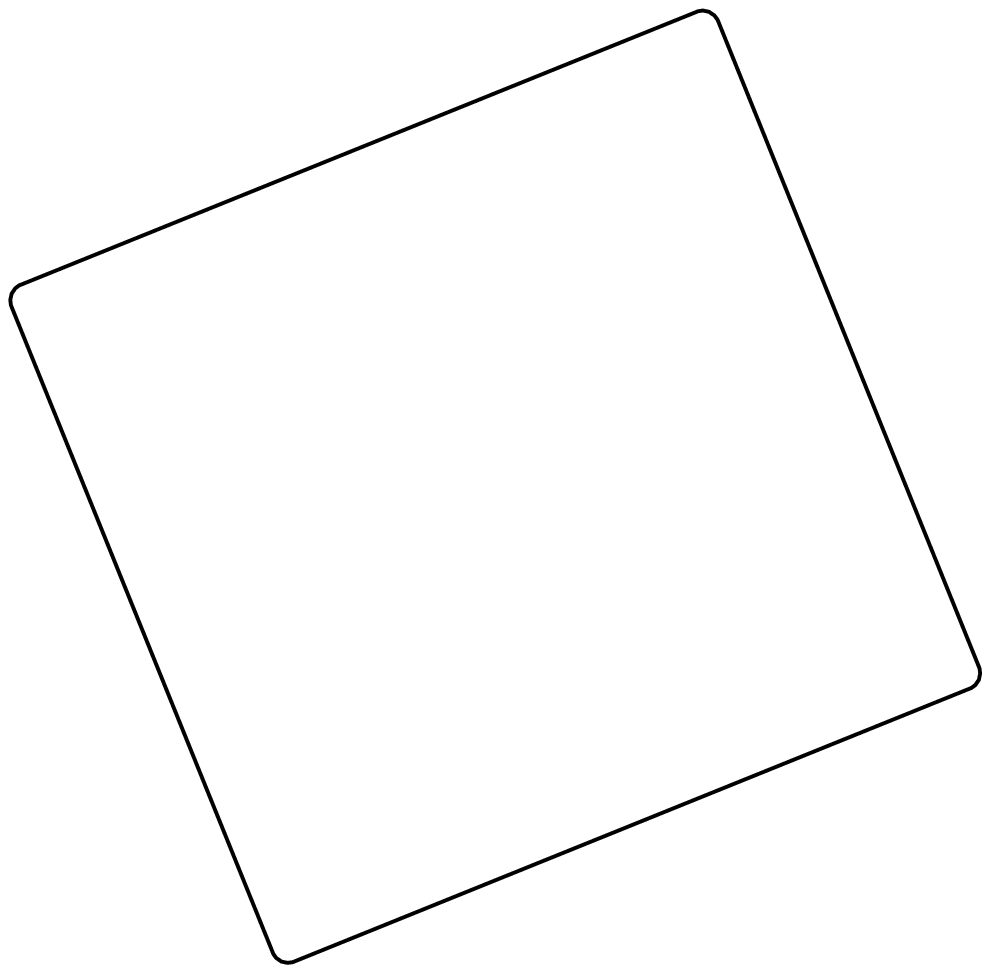} &  $----$ && $\eps_0$  \\
	16 & \includegraphics[height=1.0cm,width=1.0cm]{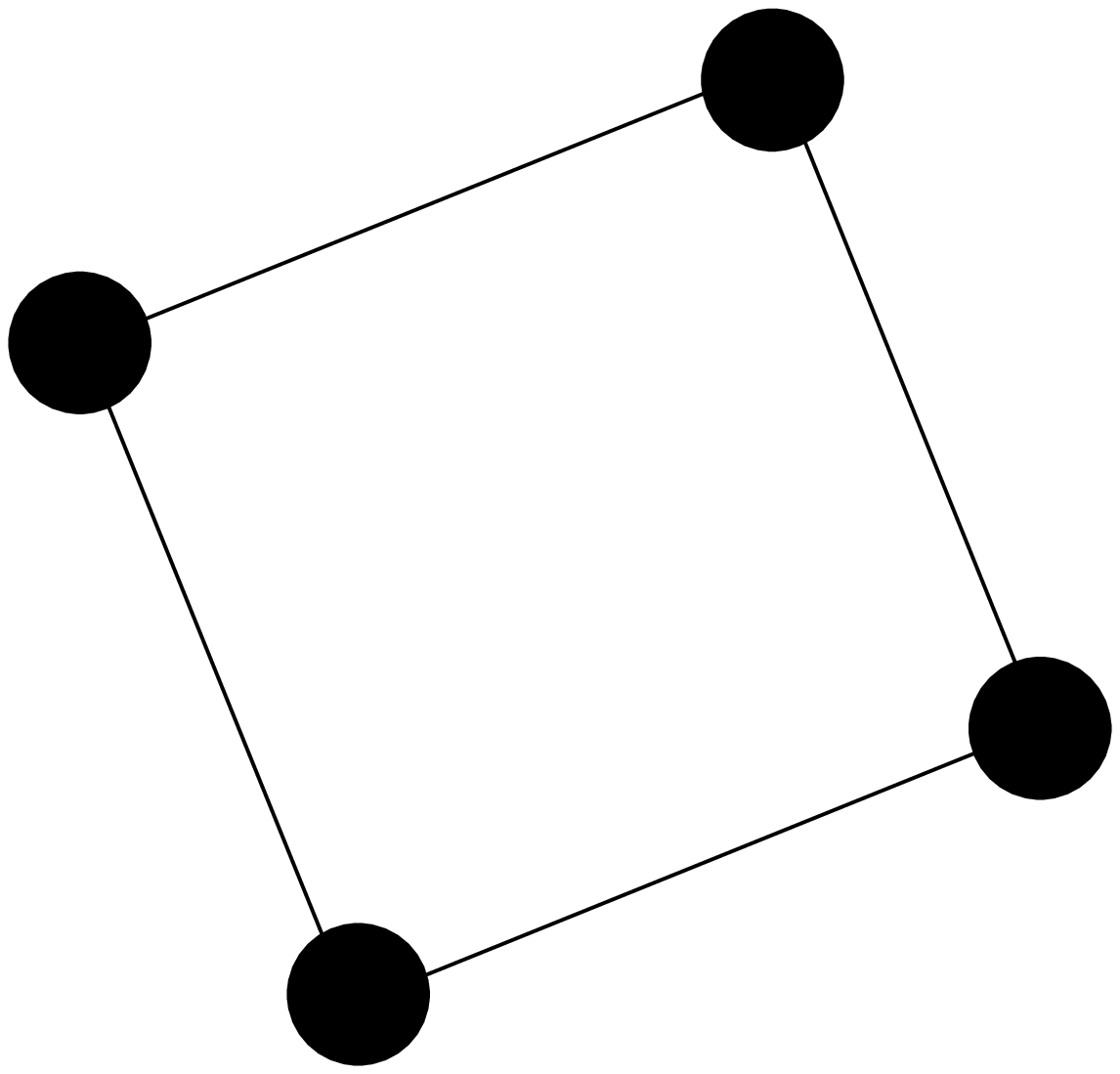} &  $++++$& & $\eps_0$  \\
	\hline
\end{tabular}

	\end{center}
\end{table}
\renewcommand{\arraystretch}{1}
\renewcommand{\tabcolsep}{1pt}

To go from the representation of proton configuration energies to the pseudospin representation we use the standard procedure \cite{blinc:66,levitskii:04,moina:20}, according to which the Hamiltonian of the
short-range correlations between protons, surrounding each A type C$_4$O$_4$ group 
is written as
\be
\label{topseudospin}
H_{yq}^A = \sum_{i=1}^{16}\hat N_i(yq){\cal E}_{i}, \quad
\hat N_i(yq)=\prod\limits_{f=1}^4\frac{1}{2}(1+s_f{\sigma}_{yqf}),
\ee
where $\hat N_i(yq)$ is the operator of the four-particle configuration $i$; $y$ stands for the layer index; $q$ is the index of the A type C$_4$O$_4$ group, and $f$ is the bond index.
$s_f=\pm1$ is the sign of the eigenvalue of the $\sigma_{yqf}$ operator in this particular configuration; ${\cal E}_{i}$ is the energy of the configuration. It is assumed
that $s_f=+1$ if the proton at the $f$th bond is localized at the H-site proximal to the given A type C$_4$O$_4$ group, and $s_f=-1$ if the proton is localized at the other (distal) H-site of the same bond.  Using Eq. (\ref{topseudospin}) and the energies ${\cal E}_{i}$ given in Table~\ref{configurations_table}, we arrive at the following expression 
for the short-range Hamiltonian, identical to that, used in the proton ordering model for the NH$_4$H$_2$PO$_4$ type antiferroelectrics (see  e.g. \cite{stasyuk:99})
\bea
&&
H_{yq}^A=V\left[\frac{\sigma_{yq1}}2\frac{\sigma_{yq2}}2+
\frac{\sigma_{yq2}}2\frac{\sigma_{yq3}}2+
\frac{\sigma_{yq3}}2\frac{\sigma_{yq4}}2+
\frac{\sigma_{yq4}}2\frac{\sigma_{yq1}}2\right]\nonumber\\
\label {H_short}&&  
 {} +   U\left[\frac{\sigma_{yq1}}2\frac{\sigma_{yq3}}2+
\frac{\sigma_{yq2}}2\frac{\sigma_{yq4}}2\right]
+\Phi \frac{\sigma_{yq1}}2\frac{\sigma_{yq2}}2\frac{\sigma_{yq3}}2\frac{\sigma_{yq4}}2\\
&&{} - \sum_{f=1}^4\left[\mu_{f1}E_1+\mu_{f3}E_3\right]\frac{\sigma_{yqf}}2.\nonumber
\eea
Here the short-range interaction constants
\be
\label{Slater-ham} V=-\frac{\eps-w_{1}}{2},\quad
U=\frac{\eps+w_{1}}{2},\quad \Phi=2\varepsilon-8w+2w_{1}
\ee
are linear functions of the Slater-Takagi type energy parameters
\[
\eps=\eps_s-\eps_a, \quad w=\eps_1-\eps_a, \quad w_1=\eps_0-\eps_a.
\]
Finally, the dipole moments $\mu_{f1}$ and $\mu_{f3}$ are the following
\bea
&&
\mu_{11}=-\mu_{31}=\mu^H+\mu^\pi_\parallel-\mu^\pi_\perp, \nonumber  \\
&&\mu_{21}=-\mu_{41}=\mu^H+\mu^\pi_\parallel+\mu^\pi_\perp, 
\eea
and
\bea
&&
\mu_{13}=-\mu_{33}=-\mu^H-\mu^\pi_\parallel-\mu^\pi_\perp, \nonumber \\ 
&& \mu_{23}=-\mu_{43}=\mu^H+\mu^\pi_\parallel-\mu^\pi_\perp.
\eea

It can be shown that just like for the KH$_2$PO$_4$  familty crystals the contributions of the correlations of the protons surrounding the A and B type groups
to the total thermodynamic potential are equal. The total Hamiltonian of the short-range interactions in this case can be written as
\begin{equation}H_{\textrm{short}} \rightarrow 2\sum_{qy}H^A_{qy}, \end{equation}
where the expression for $H^A_{qy}$ is given by Eq.~(\ref{H_short}).

The long-range interactions in the system Hamiltonian (\ref{sqa-Ham}) are the dipole-dipole interactions, in analogy to the case of KH$_2$PO$_4$ \cite{blinc:66},  renormalized by the proton-lattice coupling. The mean field approximation is believed to be an adequate approach for these interactions. In this approximation we obtain the following expressions \cite{moina:20} for the long-range intralayer 
\begin{eqnarray}\label{intra}
&& H_{\textrm{long}}^{\textrm{intra}}=-\frac12\sum_{y=1}^{N_y}\sum_{qq' \atop ff'}
J^{\textrm{intra}}_{ff'}(qq')\frac{\sigma_{yqf}}{2}\frac{\sigma_{yq'f'}}{2}\simeq  \nonumber \\
&&-2\sum_{yqf}F_{yqf}^{\textrm{intra}}\frac{\sigma_{yqf}}{2}+\sum_{yqf}F_{yqf}^{\textrm{intra}}\frac{\langle\sigma_{yqf}\rangle}{2}.
\end{eqnarray}
and interlayer 
\begin{eqnarray}
&& H_{\textrm{long}}^{\textrm{inter}}=-\frac12\sum_{y}\sum_{y'\neq y}\sum_{qq' \atop ff'}
J^{\textrm{inter}}_{ff'}(yy';qq')\frac{\sigma_{yqf}}{2}\frac{\sigma_{y'q'f'}}{2}\simeq \nonumber \\
&&
\label{inter}
-2\sum_{yqf}F_{yqf}^{\textrm{inter}}\frac{\sigma_{yqf}}{2}+\sum_{yqf}F_{yqf}^{\textrm{inter}}\frac{\langle\sigma_{yqf}\rangle}{2}
\end{eqnarray}
interactions.
Here  $N_y$ is the total number of the layers. The internal mean fields are
\begin{eqnarray}
F_{yqf}^{\textrm{intra}}=\frac 14\sum_{q'f'}J^{\textrm{intra}}_{ff'}(qq')\langle \sigma_{yq'f'} \rangle,\nonumber\\
 F_{yqf}^{\textrm{inter}}=\frac 14\sum_{y'q'f'}J^{\textrm{intra}}_{ff'}(yy;qq')\langle \sigma_{y'q'f'} \rangle. \label{fields1}
\end{eqnarray}

The four-particle cluster approximation will be used for the short-range interactions, described by the Hamiltonian (\ref{H_short}). The thermodynamic potential of the system can be written as \cite{moina:20}
\begin{eqnarray}
&& \label{tpot} G=  NU_{\textrm{seed}} +\sum_{yqf}\left(F_{yqf}^{\textrm{intra}}+F_{yqf}^{\textrm{inter}}\right)
\frac{\langle\sigma_{yqf}\rangle}{2}\\
&&
\nonumber
{} -\frac1\beta\sum_{qy}\left[2\ln \textrm{Sp} \exp (-\beta H_{qy}^{(4)}) -\sum_{f=1}^4 \ln \textrm{Sp} \exp (-\beta H_{qyf}^{(1)})\right]
,
\end{eqnarray}
where $\beta=(k_{\textrm{B}}T)^{-1}$. The four-particle cluster Hamiltonian is
\be
\label{H4}
H_{qy}^{(4)}=H^A_{qy}-\sum_{f=1}^4\frac{z_{yqf}}{\beta}\frac{\sigma_{yqf}}2,
\ee
where
\[
z_{yqf}=\beta[\Delta_{yqf}+2F_{yqf}^{\textrm{intra}}+2F_{yqf}^{\textrm{inter}}+{\mu_{f1}E_1}+{\mu_{f3}E_3}].
\]
 The fields $\Delta_{yqf}$ are the effective cluster fields that describe short-range interactions of the spin $\sigma_{yqf}$ with the particles from outside the cluster $q$. They are determined from the self-consistency condition that pseudospin mean values calculated with the four-particle  (\ref{H4}) and with the one-particle
\[ H^{(1)}_{yqf}=-\left[\Delta_{yqf}+z_{yqf}\right]\frac{\sigma_{yqf}}2\] 
Hamiltonians must coincide.

The following symmetry of the pseudospin mean values was assumed for the antiferroelectrically ordered two-sublattice model in absence of external electric field \cite{moina:20}
\begin{equation}
\label{symmetry0}
\langle \sigma_{yqf} \rangle=\exp[i{\bf k}_2 {\bf R}_y] \eta_{f},
\end{equation}
Here ${\bf k}_2=(0,b_2,0)$; $b_2$ is the basis vector of
the reciprocal lattice; the factor $\exp[i{\bf k}_2 {\bf R}_y]=\pm 1$ denotes two sublattices of an antiferroelectric;
${\bf R}_y$ is the position vector of the $y$-th layer. 
We shall call the two sublattices of the model, where $\exp[i{\bf k}_2 {\bf R}_y]=1$ and $-1$, as plus and minus sublattices, respectively. 
Also 
\begin{equation}\label{symmetry4}
\eta\equiv\eta_{1}=\eta_2=-\eta_3=-\eta_4.
\end{equation}
In fact, the system behavior in absence of the field is described by a single order parameter $\eta$.

The external electric fields $E_1$ or $E_3$ change the symmetry of the pseudospin mean values. The assumed equivalence of the hydrogen bonds linking the C$_4$O$_4$ groups along the $a$ and $c$ axes, reflected in Eq. (\ref{symmetry4}), no longer holds. Also the pseudospin mean values  in the plus and minus sublattices do not simply have the opposite signs, as in Eq.~(\ref{symmetry0}), but also different absolute values. This leads to four independent order parameters, two for the plus and two for the minus sublattices
\begin{equation}
\label{symmetry1}
\eta_{1\pm}=-\eta_{3\pm}; \quad \eta_{2\pm}=-\eta_{4\pm}.
\end{equation}

Taking into account Eq.~(\ref{symmetry1}), we can write the MFA Hamiltonians of the long-range interactions (\ref{intra}) and (\ref{inter}) as
\begin{widetext}
	\begin{eqnarray}
	\label{Hlong}
	&& H^{\textrm{intra}}_{\textrm{long}}+
	H^{\textrm{inter}}_{\textrm{long}}
	=N\left[\nu\frac{(\eta_{1+}-\eta_{1-})^2}4+
	\nu\frac{(\eta_{2+}-\eta_{2-})^2}4+
	\nu'\frac{(\eta_{1+}+\eta_{1-})^2}4+
	\nu'\frac{(\eta_{2+}+\eta_{2-})^2}4\right]\nonumber\\
	&& -2\sum_{yq}\left[
	\frac{\sigma_{qy1}-\sigma_{qy3}}{2}\left(
	\nu\frac{\eta_{1+}-\eta_{1-}}2+
	\nu'\frac{\eta_{1+}+\eta_{1-}}2
	\right)+
	\frac{\sigma_{qy2}-\sigma_{qy4}}{2}
	\left(
	\nu\frac{\eta_{2+}-\eta_{2-}}2+
	\nu'\frac{\eta_{2+}+\eta_{2-}}2
	\right)
	\right],
	\end{eqnarray}
\end{widetext}
The resulting interaction constants $\nu$ and $\nu'$
are linear combinations of the eigenvalues 
\begin{eqnarray*}
	&& \nu=\frac{J_{11}^{\textrm{intra}}(0)-J_{13}^{\textrm{intra}}(0)}4+
	\frac{J_{11}^{\textrm{inter}}({\bf k}_2)-J_{13}^{\textrm{inter}}({\bf k}_2)}4, \nonumber\\
	&& \nu'=\frac{J_{11}^{\textrm{intra}}(0)-J_{13}^{\textrm{intra}}(0)}4+
	\frac{J_{11}^{\textrm{inter}}(0)-J_{13}^{\textrm{inter}}(0)}4
\end{eqnarray*}
of the matrix of the long-range interaction constant Fourier transforms at the center of the Brillouin zone and at ${\bf k}_2$
\begin{eqnarray}
&&J^{\textrm{intra}}_{ff'}(0)=\sum_{q'}J^{\textrm{intra}}_{ff'}(qq'),\nonumber\\
&&J^{\textrm{inter}}_{ff'}(0)=\sum_{q-q'}\sum_{y-y'}J^{\textrm{inter}}_{ff'}(yy';qq'),\\
&& J^{\textrm{inter}}_{ff'}({\bf k}_2)=\sum_{q-q'}\sum_{y-y'}J^{\textrm{inter}}_{ff'}(yy';qq')\exp[i{\bf k}_2 ({\bf R}_y-{\bf R}_{y'})].\nonumber
\end{eqnarray}
We also took into account the 
the symmetry of the Fourier transforms over the bond indices $f$ and $f'$
\begin{eqnarray*}
	J_{11}=J_{22}=J_{33}=J_{44},\\
	J_{12}=J_{23}=J_{34}=J_{41},\\ J_{13}=J_{24}.
\end{eqnarray*}

Eventually, the thermodynamic potential per one
 unit cell is obtained in the following form
 \begin{widetext}
\begin{eqnarray}\label{sqa:tpot}
&& g=U_{\textrm{seed}} -\frac1\beta[\ln D_+ +\ln D_-] -\frac1{2\beta}\left[\ln(1-\eta^2_{1+}) +\ln(1-\eta^2_{2+}) +
\ln(1-\eta^2_{1-})+
\ln(1-\eta^2_{2-})\right] \\
&& {}+ \nu\frac{(\eta_{1+}-\eta_{1-})^2}4+
 \nu\frac{(\eta_{2+}-\eta_{2-})^2}4+
\nu'\frac{(\eta_{1+}+\eta_{1-})^2}4+
\nu'\frac{(\eta_{2+}+\eta_{2-})^2}4,\nonumber
\end{eqnarray}
where
\begin{eqnarray*}
	& D_{\pm}=a+\cosh(z_{1\pm}+z_{2\pm})+2b(\cosh z_{1\pm}+\cosh z_{2\pm})+\cosh(z_{1\pm}-z_{2\pm}),\\
	\label{z}
	& z_{f\pm}=\frac12\ln\frac{1+\eta_{f\pm}}{1-\eta_{f\pm}}\pm\beta\nu\frac{\eta_{f+}-\eta_{f-}}{2} + \beta\nu'\frac{\eta_{f+}+\eta_{f-}}{2}
	+
	\beta\frac{\mu_{f1}E_1}{2}+\beta\frac{\mu_{f3}E_3}{2}; \\
	& a=\exp(-\beta\eps), \quad b=\exp(-\beta w).
\end{eqnarray*}
\end{widetext}

The short-range Slater-Takagi energies, according to the model \cite{moina:20}, are considered to be quadratic functions of the H-site distance $\delta$.  In its turn, the distance $\delta$ is  taken to vary according to its experimentally observed above the transition linear temperature \cite{semmingsen:95}  dependence
\begin{equation}
\label{delta-model}
\delta=\delta_0[1+\delta_T(T-T_{\textrm{N0}})],
\end{equation}
where $T_{\textrm{N0}}$ is the transition temperature at zero electric field. It yields
\begin{equation}
\label{kdp-Slater}
\eps=\eps_0[1+\delta_T(T-T_{\textrm{N0}})]^2, \quad w=w_0[1+\delta_T(T-T_{\textrm{N0}})]^2.
\end{equation}
For the parameter of the long-range (dipole-dipole) interactions $\nu$ both the dependence of the dipole moments on $\delta$ and the changes in the interaction parameter due to the overall crystal deformation and associated with changes in the equilibrium distances between protons (dipoles) are taken into account \cite{moina:20}
\begin{eqnarray}
&\nu=\nu_0[1+\delta_T(T-T_{\textrm{N0}})]^2+\sum_{i=1}^3\psi_i\eps_i,\nonumber\\
&
\label{kdp-long} \nu'=\nu'_0[1+\delta_T(T-T_{\textrm{N0}})]^2+\sum_{i=1}^3\psi_i'\eps_i.
\end{eqnarray}

Minimization of the thermodynamic potential (\ref{sqa:tpot}) with respect to the order parameters $\eta_{f\pm}$ ($f=1,2$) and strains $\eps_i$ yields the following equations
\begin{eqnarray}
	&&\eta_{f\pm}=\frac{m_{f\pm}}{D_\pm},\\
	&&\label{eq:strains}
	0=\sum_{j=1}^3c_{ij}^{(0)}[\eps_j-\alpha_j^{(0)}(T-T_j^0)]\\
	&&\qquad+
\sum_{f=1}^2\left\{\psi_i\frac{\eta_{f+}-\eta_{f-}}{2v}\left[\frac{\eta_{f+}}2-\frac{\eta_{f-}}{2}-\frac{m_{f+}}{D_+}+\frac{m_{f-}}{D_-}\right]\right.\nonumber\\
&&\qquad+
\left.\psi'_i\frac{\eta_{f+}+\eta_{f-}}{2v}\left[\frac{\eta_{f+}}2+\frac{\eta_{f-}}{2}-\frac{m_{f+}}{D_+}-\frac{m_{f-}}{D_-}\right]\right\},\nonumber
\end{eqnarray}
where
\[m_{f\pm}=\sinh (z_{1\pm}+z_{2\pm})+2b\sinh z_{f\pm}\pm \sinh (z_{1\pm}-z_{2\pm}).\]
The anomalous parts of the strains (the second sum in (\ref{eq:strains})) are quadratic functions of the order parameters $\eta_{f\pm}$, indicating the electrostrictive origin of these parts.

The net crystal polarizations $P_1$ and $P_3$ are obtained as
\begin{equation}
\label{p1}
P_j=-\frac 1v\frac{\partial g}{\partial E_{j}}=P_{j+}+P_{j-},
\end{equation}
where
\begin{eqnarray}
&& P_{1\pm} = 
\frac{\mu^H+\mu^\pi_\parallel}{2v}
(\eta_{1\pm}+\eta_{2\pm}) + \frac{\mu_\perp^\pi}{2v}(\eta_{2\pm}-\eta_{1\pm}),\nonumber \\
&& P_{3\pm}=
\frac{\mu^H+\mu^\pi_\parallel}{2v}
(\eta_{2\pm}-\eta_{1\pm}) - \frac{\mu_\perp^\pi}{2v}(\eta_{2\pm}+\eta_{1\pm})
\label{net}
\end{eqnarray}
are the sublattice polarizations. The total net polarization within the $ac$ plane is
\begin{equation}
P=\sqrt{(P_{1+}+P_{1-})^2+(P_{3+}+P_{3-})^2}.
\end{equation}

The static  dielectric permittivity $\eps_{11}$ can be  found by numerical differentiation of the net polarization $P_1$ (\ref{p1}) with respect to the electric field $E_1$ 
\[\eps_{11}=\frac{1}{\eps_0}\left(\frac{\partial P_1}{\partial E_1}\right)_\sigma,\]
where $\eps_0\approx 8.85$~F/m is the permittivity of free space. The differentiation is carried out at constant external stresses $\sigma_i$ (equal to zero in our case), i.e. the permittivity of a mechanically free crystal is calculated.

\section{Calculations}
\label{calculations}
Even though the developed here model allows us to explore the influence of an electric field of any arbitrary orientation within the $ac$ plane, we restrict our numerical calculations by the  field $E_1$ only. 

\begin{table*}[htb]
	\caption{The adopted values of the model parameters. All, except $\mu$ and $\nu_0'$, are taken from \cite{moina:20}. }
	\label{tbl1}
	\begin{center}
		\renewcommand{\arraystretch}{0}
		
		\begin{tabular}{ccccccc|cccc|ccc|c|cc}
			\hline
			$\eps_0/k_{B}$ & $w_0/k_{B}$ &  $\nu_0/k_{B}$ &  $\nu'_0/k_{B}$ & $\psi_1/k_{B}$ & $\psi_2/k_{B}$ & $\psi_3/k_{B}$ &		$c_{11}^{0}$&  $c_{12}^{0}$ & $c_{13}^{0}$ & $c_{22}^{0}$ & $\alpha_1^0$ & $\alpha_2^0$ &$\delta_T$  & $v$ & $\mu^H+\mu^\pi_\parallel$ & $\mu^\pi_\perp$  \\
			\multicolumn{7}{c|}{(K)}  &			\multicolumn{4}{c|}{($10^{10}$ N/m$^{2}$)} & \multicolumn{3}{c|}{($10^{-5}$ K$^{-1}$)}  & ($10^{-28}$m$^3$) 
			& \multicolumn{2}{c}{($10^{-29}$~C m)} \\
			\hline
			395 &  1100 & 79.8 & -50 &  -518
			& 445 & 1096 &	6.5 & 2.3 & -3.1 & $2.38-0.02T$ & 1.2 & 13.0 & 20 & 2.0 &3.64 &1.65  \strut 	\\ 	\hline
		\end{tabular}
		\vspace{1ex}
		\renewcommand{\arraystretch}{1}
	\end{center}
\end{table*}

The thermodynamic potential is minimized numerically
with respect to the four order parameters $\eta_{f\pm}$. Simultaneously, the strains $\eps_i$ are determined from Eqs.~(\ref{eq:strains}).

The values of most of the model parameters have been determined earlier \cite{moina:20}. They were required to provide the best fit to the experimental temperature curves  of
the sublattice polarization (order parameter)  at ambient pressure, temperature and hydrostatic pressure dependences of the lattice strains  $\eps_i$, and the pressure dependence of  the transition temperature $T_{\textrm N}$.  Details of the fitting procedure are given in \cite{moina:20}.

\begin{figure}[hbt]
	\centerline{\includegraphics[width=\columnwidth]{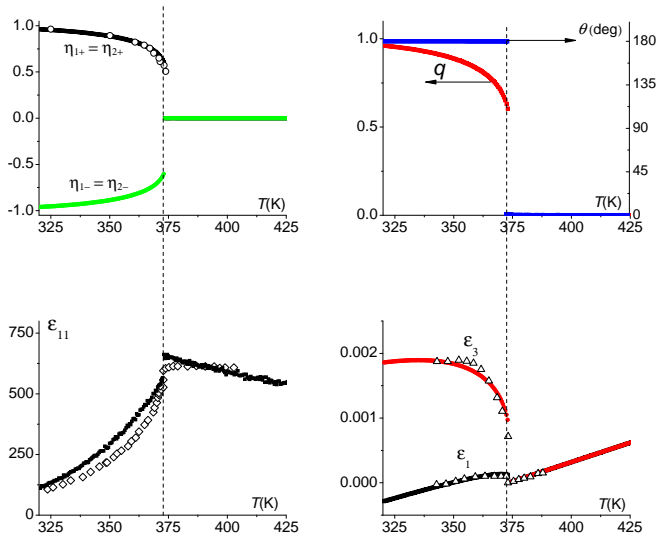}}
	\caption{The temperature dependences of the order parameters $\eta_{f\pm}$ (upper left), AFE ordering parameter $q$ and non-collinearity anlge $\theta$ (upper right), dielectric permittivity $\eps_{11}$ (lower left), and lattice strains $\eps_1$ and  $\eps_3$ (lower right) of squaric acid at $E_1=0$. Open symbols: experimental points, taken from Refs. \cite{klymachyov:97} ($\circ$),  \cite{horiuchi:18} ($\diamond$),  \cite{ehses:81} ($\bigtriangleup$). Vertical dashed lines are a guide to the eye, showing the phase transition. } \label{fig-e0}
\end{figure}

New to the present model are the parameters $\nu'(0)$, $\psi'$ (see Eq.~(\ref{kdp-long})) and the  dipole moments $\mu^H+\mu_\parallel^\pi$ and $\mu_\perp^\pi$.
For the sake of simplicity we assume that $\psi'_i=\psi_i$.  The parameter $\nu'(0)$ and the dipole moments are determined by fitting the calculated dielectric permittivity $\eps_{11}$ curve at zero external bias field to the experimental points of Ref.~\cite{horiuchi:18}. It cannot be said, however, that the adopted set of $\nu'(0)$ and $\mu$ is unique.
The obtained agreement with the experimental data is illustrated in fig.~\ref{fig-e0}. The used values of the model parameters are summarized in Table~\ref{tbl1}.

For the analysis of the field-induced phase transitions in the system it is convenient to introduce some new  parameters. First, those are the parameters
of ferroelectric and antiferroelectric ordering $Q$ and $q$, which are, basically, the sum and the difference of proton polarizations of the two sublattices, respectively
\begin{eqnarray}
\label{Qfe}
Q=\frac14\left(\eta_{1+}+\eta_{2+}+\eta_{1-}+\eta_{2-} \right),\\
\label{qafe}
q=\frac14\left(\eta_{1+}+\eta_{2+}-\eta_{1-}-\eta_{2-} \right).
\end{eqnarray}

But perhaps most important is the new parameter
$\theta$, the non-collinearity angle, i.e. the angle between polarization vectors of the plus and minus sublattices
\begin{equation}
\label{theta}
\theta=\left|\arccos\frac{\eta_{1+}+\eta_{2+}}{\sqrt{2(\eta_{1+}^2+\eta_{2+}^2)}}-\arccos\frac{\eta_{1-}+\eta_{2-}}{\sqrt{2(\eta_{1-}^2+\eta_{2-}^2)}}\right|.
\end{equation}
Each of the terms here is the angle between the actual plus/minus sublattice polarization vector, which components are given by Eqs.~(\ref{net}), and the polarization vector of the fully ordered plus sublattice, determined from Eqs.~(\ref{net}) at $\eta_{f+}=1$. It should be noted that the non-collinearity angle does not depend on the dipole
moments $\mu^H+\mu^\pi_\parallel$ or $\mu^\pi_\perp$ explicitly.

Under the positive external field, it is the polarization in the minus sublattice that is switched, whereas the plus sublattice polarization  stays basically constant. The angle of the minus polarization rotation is then close to $\pi-\theta$.

Orientation of the sublattice polarization vector is determined by relative populations of four lateral proton configurations in it (configurations 1-4 in Table~\ref{configurations_table}). In our model these populations can be deduced from the pseudospin mean values
$\eta_{1\pm}$ and $\eta_{2\pm}$. Thus, $\eta_{1-} \approx \eta_{2-}\approx - 1$ means
that almost all C$_4$O$_4$ groups in the minus sublattice are in the configuration 3; this is the assumed ground state configuration in absence of the electric field. Then
$\eta_{1-} \approx -\eta_{2-}\approx - 1$ means the configuration 2 and
a 90$^\circ$ rotation of the minus sublattice polarization ($\theta=\pi/2$). $\eta_{2-}\sim 0$, would mean a combination of equally populated configurations 2 and 3, or a statistically averaged 45$^\circ$  rotation of the  polarization  of the sublattice ($\theta=3\pi/4$), and so on.

One should expect the presence of the new phase with the 90$^\circ$ rotated polarization of the minus sublattice in the $T-E_1$ phase diagram. Horiuchi \textit{et al} \cite{horiuchi:18} call the latter phase as the ferroelectric FE-$\alpha$ phase. Strictly speaking, it is not ferroelectric, but non-collinear ferrielectric, with perpendicular and different by their magnitude polarizations of the two sublattices. We shall denote this phase as NC90. The predicted by calculations \cite{horiuchi:18} FE-$\beta$ phase with parallel polarizations of the two sublattices is indeed collinear ferroelectric; it is FE in our notations. 

%

Figure~\ref{fig-e0} shows that the system behavior in zero electric field is of a typical antiferroelectric with the first order phase transition between the AFE and PE phases. The AFE ordering parameter $q$ is identical to the pseudospin mean value in the plus sublattice; the parameter of the FE ordering $Q$ is zero at any temperature, and the non-collinearity angle $\theta$ is 180$^\circ$ in the AFE phase and 0 above the transition.

To construct the $T-E_1$ phase diagram of the squaric acid and detect the phase transitions in the system we analyzed the temperature behavior of the quantities from fig.~\ref{fig-e0} at different values of the external field $E_1$, as well as the field curves of total net polarization $P$. The temperature curves of these quantities at different fields can be found in Appendix.

The obtained phase diagram is shown in fig.~\ref{pd}.
As one can see, we detected three lines of the first order phase transitions I, II, and III, each terminating at the critical point CP$_1$, CP$_2$, CP$_3$, respectively, and a line of the second order phase transitions IV, terminating at two critical end points CEP$_1$ and CEP$_2$ at the lines of the first order transitions I and III.

\begin{figure}[hbt]
	\centerline{\includegraphics[width=\columnwidth]{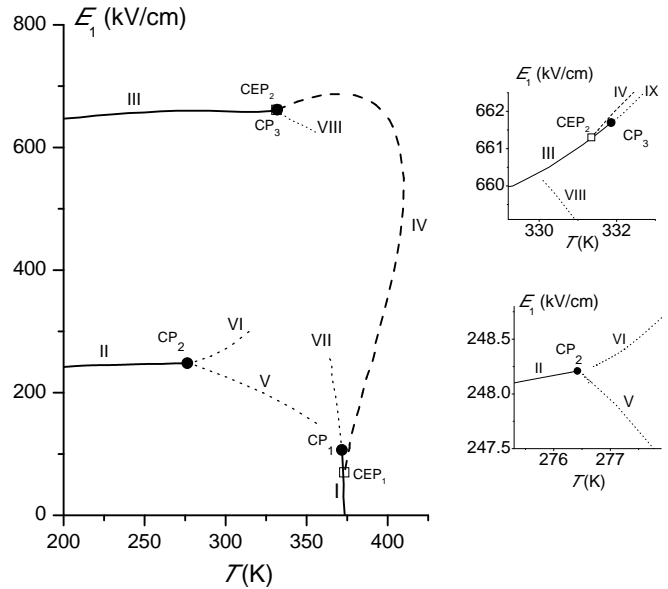}}
	\caption{The $T-E_1$ phase diagram of the squaric acid. Solid and dashed lines indicate the first and second order phase transitions, respectively. Dotted lines are the supercritical lines, corresponding to smeared anomalies in the temperature curves of the dielectric permittivity. Open squares and full circles indicate the  critical end points and critical points, respectively. Right panel: closeups of the phase diagram in the vicinities of CP$_2$ and CP$_3$.} \label{pd}
\end{figure}

Positions (the fields, mostly) of the upper phase transition line  III, as well of the critical end point CEP$_2$ and critical point CP$_3$, essentially depend on our choice of the values of the model parameters $\mu^H+\mu^\pi_\parallel$ and $\mu^\pi_\perp$.
The topology of the phase diagram, however, remains the same, and the positions of experimentally attainable lines I, II, and IV, as well as of the points CEP$_1$, CP$_1$, and CP$_2$, are much less sensitive to reasonable variations of $\mu^H+\mu_\parallel$ and $\mu_\perp$.
At the present choice of the model parameters the coordinates of the 
critical points and critical end points are the following

\begin{tabular}{lll}
CP$_1$:& $T_{\textrm{cp}1}=371.8$~K, &	$E_{\textrm{cp}1}=106.5$~kV/cm, \\
CP$_2$: & $T_{\textrm{cp}2}=276.3$~K,	&$E_{\textrm{cp}2}=248.2$~kV/cm, \\
CP$_3$: & $T_{\textrm{cp}3}=331.9$~K,	&$E{\textrm{cp}3}=661.7$~kV/cm,\\
CEP$_1$:& $T_{\textrm{cep}1}=373.1$~K, 	&$E_{\textrm{cep}1}=49.5$~kV/cm,\\
CEP$_2$:& $T_{\textrm{cep}2}=331.4$~K, 	&$E_{\textrm{cep}1}=661.3$~kV/cm.\\
\end{tabular}

In addition to the sharp anomalies at the first and second order phase transitions, 
the temperature curves of the dielectric permittivity also exhibit several smeared anomalies at different values of the field (see, e.g. figs.~\ref{fig-e225}-\ref{fig-e658} in Appendix). 
Loci of these anomalies are indicated in the phase diagram by the dotted lines V to IX. The lines V, VI, VIII, and IX correspond to rounded maxima and the line VII to rounded cusps in the temperature curves of the permittivity. Some of these dotted lines appear to be continuations of the lines of the first order transitions I, II, and III beyond the critical points CP$_1$, CP$_2$, and CP$_3$, respectively, while the line VIII terminates at the line III of the first order transitions just below the critical point CP$_3$ (see the inset to fig.~\ref{pd}). The further from the critical points, the more smeared and less pronounced are the permittivity anomalies. These dotted lines end, when the anomalies become too smeared to be discernible. 

The lines, emanating from the critical points are, apparently, analogs of the Widom lines, observed in ferroelectrics in electric fields applied along the axis of spontaneous polarization \cite{kutnjak:07,novak:13,ma:16}, while the line VIII is obviously  not. In any case, classification of the found supercritical lines lies beyond the scope of the present study.

\begin{figure}[hbt]
	\centerline{\includegraphics[width=\columnwidth]{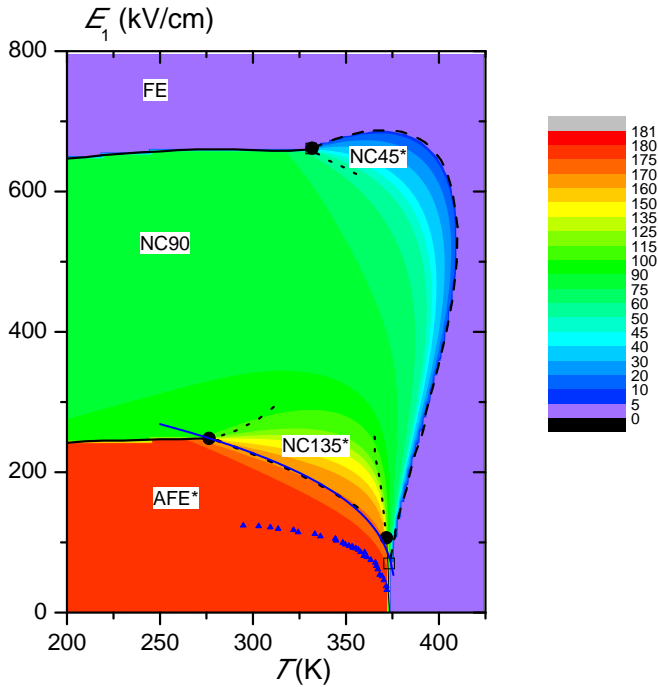}}
	\caption{The $T-E_1$ phase diagram of the squaric acid, overlapping the color gradient $T-E_1$ plot of the non-collinearity angle $\theta$. Black lines and symbols are the same as in fig.~\ref{pd}.  Blue full triangles $\blacktriangle$ are the experimental points \cite{horiuchi:18} for the transition to the FE-$\alpha$ phase; solid blue line: $E_1=53.5(T_f-T)^{1/3}$ (see text).  } \label{pdtheta}
\end{figure}

To clarify the physical meaning of the observed phase transitions and of the smeared anomalies of the permittivity, it is most efficient and informative to look at the phase diagram overlapped with the $T-E_1$ color gradient plot of the non-collinearity angle $\theta$ as in fig.~\ref{pdtheta}.

The phase, existing at low temperatures and fields (shown in red), is non-collinear ferrielectric, where a minor rotation of the minus sublattice polarization takes place, discernible mostly at its upper right boundary. It is very close to the unitial antiferroelectric phase, with nearly antiparallel sublattice polarizations and $\theta\sim 180^\circ$. We shall denote this phase as AFE*.

 On increasing the field at temperatures below  $T_{\textrm{cp}2}$ the system undergoes two successive first order phase transitions: one is accompanied by the change of $\theta$ from $\sim$180$^\circ$ to $\sim$90$^\circ$, whereas at the other $\theta$ jumps from $\sim$90$^\circ$ to $0^\circ$. This is a clear indication of 
two transitions associated with the polarization switching: the first one is from the AFE* phase to the non-collinear ferrielectric phase with the minus sublattice polarization is rotated by $90^\circ$ (NC90, the green region), whereas the second is from the NC90 phase to the shown in purple collinear ferroelectric phase (a total rotation of the minus sublattice polarization by $180^\circ$, as compared to the initial AFE phase). Thus, our theory corroborates the findings of Horiuchi et al \cite{horiuchi:18,ishibashi:18} about the existence of two field-induced phases in squaric acid.

The supercritical lines V, VI and VII, were they closed, would apparently encompass a region of the phase diagram (the orange to yellow region in fig.~\ref{pdtheta}), which by its shape resembles closely the region of the semipolar phase in uniaxial antiferroelectrics \cite{suzuki:83,suzuki:83:2}. Roughly, in our case it is the region, where the non-collinearity angle  $\theta$ changes gradually from $\sim$180$^\circ$ to $\sim$90$^\circ$ (the minus sublattice polarization gradually rotates to become perpendicular to the plus sublattice polarization). We shall denote this transitional region as NC135*. Across it the system goes from the AFE* phase to NC90 phase without undergoing any actual phase transition.  As one can see in fig.~\ref{pdtheta}, the supercritical lines, along which smeared anomalies of the permittivity occur, correlate with the onset and the end of the gradual transition from the AFE* to the NC90 phase.

The supercritical line V between the AFE* and NC135* regions is well described by the power law
$\sim(T_f-T)^{1/3}$, with $T_f=T_{\textrm{N0}}+3$~K.
Qualitatively this resembles the experimental temperature variation
$\sim(T_{\textrm{N0}}-T)^{1/3}$ 
of the  switching fields $E_{sw}$ for the transition to the FE-$\alpha$ phase between \cite{horiuchi:18}.  Quantitatively, however, the calculated transition fields are about twice higher than experimental  ones.

There is the second ``transitional'' region in the phase diagram, similar by its shape and nature to the NC135* one. Across this region the non-collinearity angle gradually changes from $\sim$90$^\circ$ to zero (the minus sublattice polarization rotates once again by $\sim$90$^\circ$ and becomes collinear to the plus sublattice polarization). It is shown in blue colors in the phase diagram and is limited partially by the second-order phase transition line IV and the supercritical line VIII. It starts from the critical end point CEP$_2$ at high fields and continues along the line IV down to the narrow wedge between IV and the first order transition line I with the tip at CEP$_1$. We shall denote this transitional region as NC45*.

The phase observed at high fields and low temperatures and the phase observed at high temperatures at any non-zero field (the purple region on the phase diagram) is the same collinear field-induced ferroelectric phase (FE) with $\theta=0$. The true paraelectric and antiferroelectric phases exist only at $E_1=0$. A transition from any non-collinear phase to the collinear FE phase can be realized only by crossing at least one line of the phase transitions of the first or second order.

\begin{figure}[hbt]
	\centerline{\includegraphics[width=0.8\columnwidth]{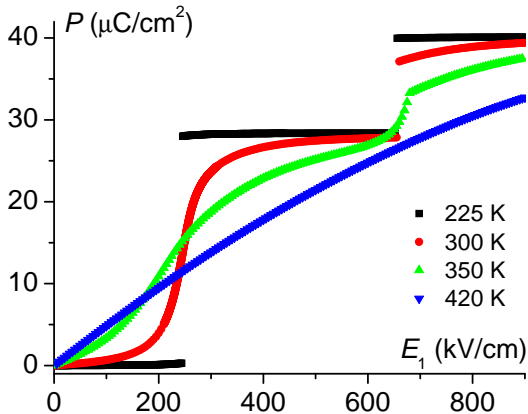}}
	\caption{The calculated field dependences of total polarization $P$  of the squaric acid at different temperatures.  } \label{fig-pol}
\end{figure}

Some characteristic (but not all possible) types of the field dependences of the total crystal net polarization $P$  at different temperatures are shown in fig.~\ref{fig-pol}. At low temperatures the system undergoes two first order phase transitions with jumps of polarization, when the lines II and III are crossed. When temperature increases, the lower jump is replaced with a steep but gradual increase
of the polarization, as the field increases across the region NC135* of the phase diagram, while the upper jump of the polarization persists. At further increase of temperature
both first order transitions are replaced with gradual increases, which correspond to going across the regions NC135* and then NC45*; also, a cusp at crossing the line of the second order transitions IV is seen. Finally, at high temperatures the system remains in the collinear FE phase at all non-zero fields, and the field dependence of $P$ is a monotonic increase. The theoretical absolute values of polarization in the fully ordered NC90 and FE phases at low temperatures are nearly twice higher than the predicted 13.6 and 23.2~$\mu$C/cm$^2$ \cite{horiuchi:18}.

Detailed illustration of the variety of possible temperature behavior
of the order parameters, non-collilnearity angle, permittivity, and lattice strains at different values of the electric field $E_1$ can be found in Appendix.

\section{Concluding remarks}
We have presented a modification to the deformable two-sublattice proton ordering model \cite{moina:20}, aimed to describe the effects of the external electric fields applied to antiferroelectric crystals of squaric acid and confined to the plane of the hydrogen bonds. The dipole moments associated directly with protons and with the $\pi$ bonds that can be switched by proton rearrangement are included into the model.

The major obtained field effect is the rotation of polarization in the minus sublattice. The system behavior is then best described by the introduced non-collinearity angle $\theta$, which is the angle between the polarizations vectors of the plus and minus sublattices. This parameter is expressed in terms of the pseudospin mean values, Eq.~(\ref{theta}). Its temperature  dependences  at different fields closely resemble those of the parameter of antiferroelectric ordering $q$, Eq.~(\ref{qafe}).

The constructed $T-E_1$ phase diagram consists of the non-collinear ferrielectric phase with nearly antiparallel sublattice polarizations (AFE*, $\theta\sim 180^\circ$), non-collinear phase with nearly perpendicular sublattice polarizations (NC90, $\theta\sim90^\circ$), and collinear ferroelectric phase (FE, $\theta=0$).
There are also two transitional regions in the diagram, where the non-collinearity angle $\theta$ changes gradually from about 180$^\circ$ to about 90$^\circ$ (region NC135*), and from about 90$^\circ$ to zero (region NC45*). These regions are not separated by any phase transition line from the AFE* or NC90 phases, but are partially encompassed by the supercritical lines, along which the smeared anomalies of the permittivity are observed. Then, rotation of the minus sublattice polarization (transition between phases), depending on the chosen path within the $E_1-T$ plane, can occur either discontinuously via a first order phase transition or continuously, by going across a transitional region.

Note that the predicted continuous rotation of the sublattice polarization is a statistically averaged phenomenon. The dipole moment of each H$_2$C$_4$O$_4$ group
can only be oriented along the four specific directions, described in Table~\ref{configurations_table}. At low temperatures, in the fully ordered system ($|\eta_{f\pm}|=1$) the polarizations of the sublattices are either collinear in the AFE* or FE phases or perpendicular in the NC90 phase, in accordance with the pseudotetragonal symmetry of the hydrogen bond network in squaric acid. However, at higher temperatures, due to the interplay of the switching external field and thermal fluctuations, protons in some of the H$_2$C$_4$O$_4$ groups can rearrange into other configurations, thereby switching the group dipole moment. Because of this disorder, the \textit{average} vector of the sublattice polarization can be directed arbitrarily, not just along
the specific directions, as in the fully ordered system at low temperatures.

The theoretical results qualitatively agree with the experimental findings by Horiuchi et al \cite{horiuchi:18}. The calculated values of the net polarization and the transition fields, however, are perceptibly higher than observed experimentally. The agreement with experiment, at least for the transition fields, could possibly be improved by inclusion of non-linear terms into the system Hamiltonian. Such calculations are currently underway and will be published elsewhere. To explore in detail the nature of the observed several supercritical lines in the temperature-electric field phase diagram would also be very interesting.

\section*{Appendix}
\appendix*

In the discussion below we use the notations for the phase transition lines from fig.~\ref{pd} and for the phases from fig.~\ref{pdtheta}.

When the electric field $E_1$ is increased until up to
$E_{\textrm{cep}1}$, the system behavior remains basically the same as at zero field (fig.~\ref{fig-e0}). The major changes are trivial: the appearance of small non-zero values of $\eta_{f\pm}$ and $Q$ above the transition. 

\begin{figure}[hbt]
	\centerline{\includegraphics[width=\columnwidth]{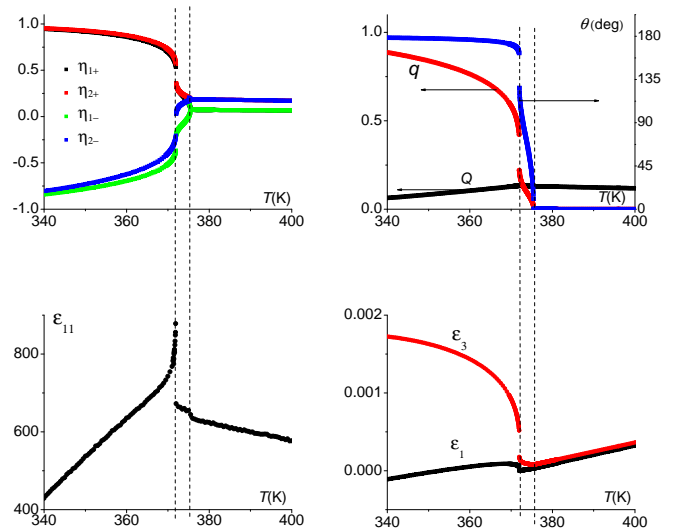}}
	\caption{The calculated temperature dependences of the order parameters $\eta_{f\pm}$ (upper left), AFE and FE ordering parameters $q$ and $Q$, the non-collinearity angle $\theta$ (upper right), dielectric permittivity (lower left), and lattice strains $\eps_1$ and  $\eps_3$ (lower right) of squaric acid at $E_1=100$~kV/cm. Vertical dashed lines are drawn through the phase transition points} \label{fig-e100}
\end{figure}

Qualitative changes occur when the field is raised above $E_{\textrm{cep}1}$. At fields between $E_{\textrm{cep}1}$
and $E_{\textrm{cp}1}$ (see fig.~\ref{fig-e100}), the system undergoes two successive  phase transitions of the first and then of the second order at crossing the lines I and IV on increasing temperature 
in the vicinity of $T_{\textrm{N0}}$. Between them the non-collinearity angle $\theta$ rapidly decreases from about 135$^\circ$ at the transition from the AFE* phase down to zero at the transition to the FE phase. At the first order transition the permittivity has a very high sharp peak, whereas  only a small jump is observed at the second order transition.
With a further increase of the field the jumps of the order parameters and pseudospin mean values at the lower transition (along the line I) decrease, until at
$E_{\textrm{cp}1}$ this transition becomes of the second order: the critical point CP$_1$ is reached.

\begin{figure}[hbt]
	\centerline{\includegraphics[width=\columnwidth]{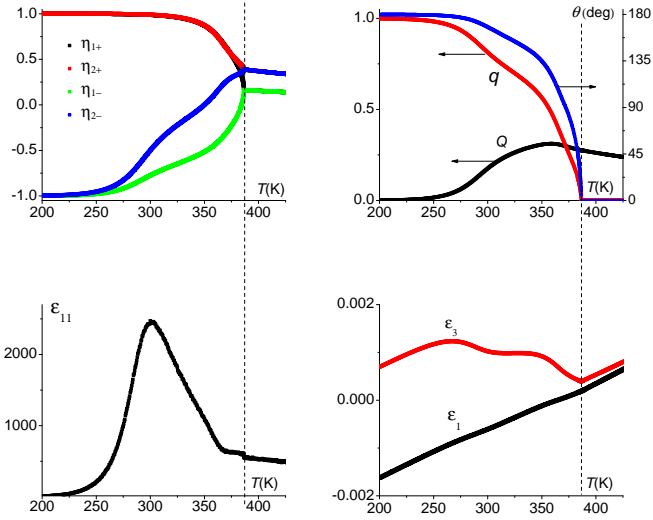}}
	\caption{Same as in fig.~\ref{fig-e100} at $E_1=225$~kV/cm. } \label{fig-e225}
\end{figure}

At  fields above $E_{\textrm{cp}1}$ the system undergoes only the second order transition at the line IV (see fig.~\ref{fig-e225}). The sharp peaks of the dielectric permittivity and jumps of the order parameters  $\eta_{f-}$, $Q$, and $q$, as well as the non-collinearity angle $\theta$ at the lower transition at the line I, observed at the fields below $E_{\textrm{cp}1}$, transform now into smeared cusps of the permittivity, forming the supercritical line VII, and inflection points in the temperature curves of the order parameters and $\theta$. Additionally, there arises another smeared anomaly at much lower temperature at crossing the supercritical line V. This anomaly appears as a high rounded peak of the permittivity, and inflections in the temperature curves of $\eta_{f-}$, $Q$, $q$, and $\theta$ are also observed. 

The temperatures between  the permittivity peak and cusp correspond to the transitional region NC135*, where the non-collinearity angle changes gradually from about 180$^\circ$ to 90$^\circ$. As one can see, the temperature curve of $\theta$ basically repeats the curve of the AFE ordering parameter $q$.

\begin{figure}[hbt]
	\centerline{\includegraphics[width=\columnwidth]{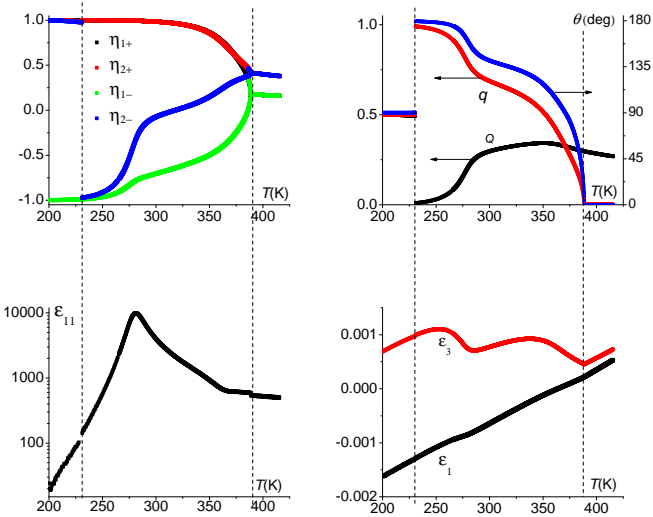}}
	\caption{Same as in fig.~\ref{fig-e100} at $E_1=245$~kV/cm. } \label{fig-e245}
\end{figure}

At higher fields there is a narrow field range just below $E_{\textrm{cp}2}$, where
one can cross the line II of the first order transitions  (see fig.~\ref{fig-e245}). At low temperatures the system is in the NC90 phase, but undergoes the first order transition to the  
 phase with increasing temperature. At this transition $\eta_{2-}$ abruptly changes its sign, and the non-collinearity angle $\theta$ jumps from 90$^\circ$ to 180$^\circ$.  Jumps of the dielectric permittivity and order parameters $Q$ and $q$ also take place. Interestingly, the strains $\eps_1$ and $\eps_3$ do not exhibit any anomaly at this transition. This is to be expected, since the anomalous parts of the strains, being of the electrostrictive origin, are proportional to the squares of the order parameters (see eqs.~(\ref{eq:strains})), so the anomalies are to be observed, when the absolute values of the order parameters are changed at the transition. At this particular transition, however, only the sign of the order parameter $\eta_{2-}$ is changed, whereas its absolute value remains basically the same. 
 
 At higher temperatures the system behavior is the same as in fig.~\ref{fig-e225}: a gradual transition through the NC135* phase, and then a second order phase transition to the FE phase. Notable are very high values of the rounded permittivity peaks at crossing the supercritical line V.

With a further increase of the field the jump of the order parameters at crossing the line II decreases, until at  $E_{\textrm{cp}2}$
the transition becomes of the second order, and the critical point CP$_2$ is reached. 

\begin{figure}[hbt]
	\centerline{\includegraphics[width=\columnwidth]{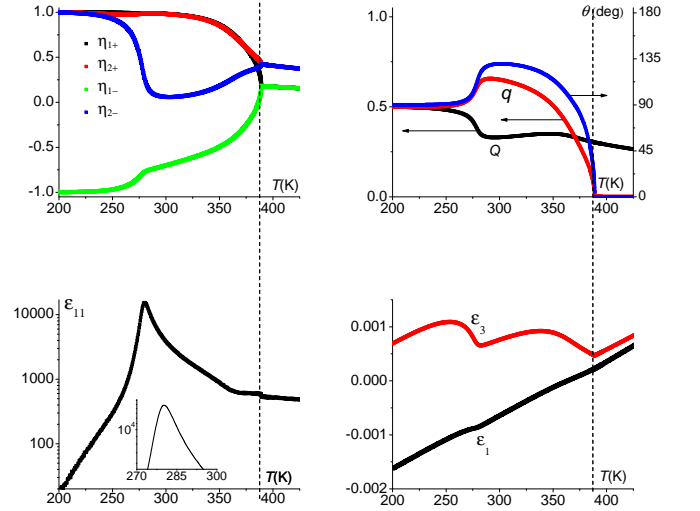}}
	\caption{Same as in fig.~\ref{fig-e100} at $E_1=250$~kV/cm. } \label{fig-e250}
\end{figure}

Above $E_{\textrm{cp}2}$ (see fig.~\ref{fig-e250}) a large smeared  anomaly of the permittivity, as well as inflections in the temperature curves of $\eta_{f-}$, $Q$, $q$, and $\theta$  still can be observed in a certain range of the field, as the system crosses the supercritical line VI at increasing temperature and enters the ``transitional'' region NC135* from the NC90 phase. The inset shows that the permittivity peak is indeed smeared out. A cusp of the permittivity is observed at crossing the supercritical line VII, when the system leaves the NC135* region. The second order phase transition to the FE phase at the line IV takes place at a further increase of temperature.

As the field $E_1$ is increased, the smeared anomalies at the supercritical lines VI and VII become less and less pronounced,
so up to about 650~kV/cm (below the line III) only small jumps of the permittivity  at the second order transition to the 
FE phase persist.

\begin{figure}[hbt]
	\centerline{\includegraphics[width=\columnwidth]{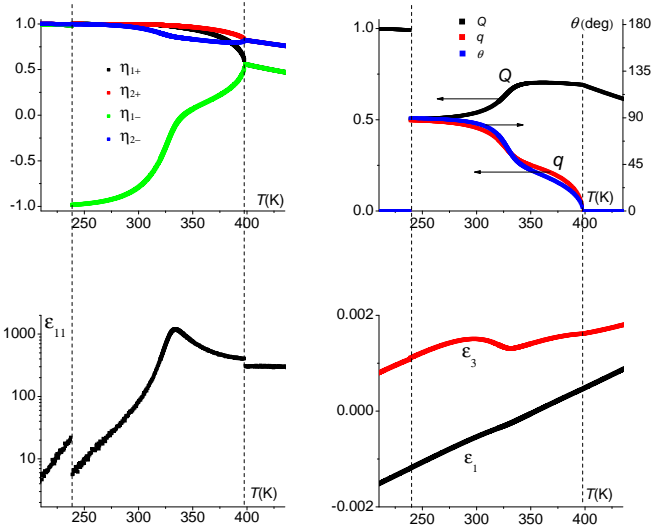}}
	\caption{Same as in fig.~\ref{fig-e100} at $E_1=655$~kV/cm. } \label{fig-e655}
\end{figure}

The situation, when the field is just high enough to cross the line III is illustrated in fig.~\ref{fig-e655}. At low temperatures both $\eta_{1-}$ and $\eta_{2-}$ are positive, and the system is in the high-field collinear FE phase. At increasing temperature it undergoes the first order phase transition back to the NC90 phase at the line III, which is reflected in the jumps of $\eta_{1-}$ to nearly $-1$, of $\theta$ from zero to 90$^\circ$, and in a small downward jump of the permittivity $\eps_{11}$.  The rounded peak of the permittivity corresponds to the supercritical line VIII and signals the start of the gradual transition to the region NC45*, in which the non-collinearity angle $\theta$ varies from 90$^\circ$ to zero. Then the system undergoes the second order phase transition to the FE phase at the line IV.

\begin{figure}[hbt]
	\centerline{\includegraphics[width=\columnwidth]{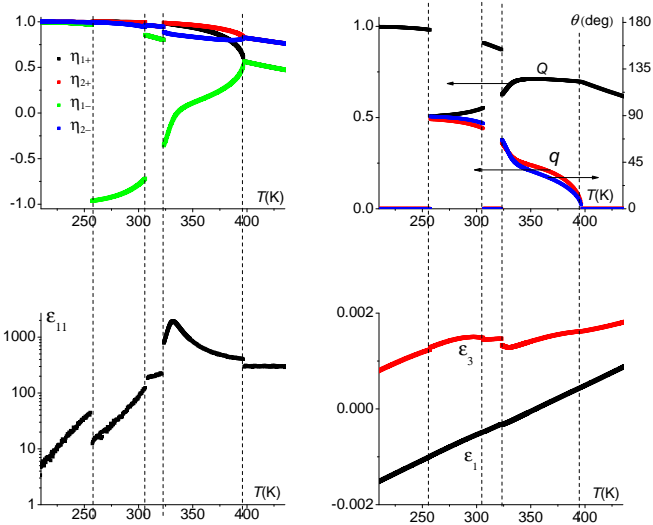}}
	\caption{Same as in fig.~\ref{fig-e100} at $E_1=658$~kV/cm. } \label{fig-e658}
\end{figure}

Since the line III of the first order phase transitions is non-monotonic, there exist a narrow range of the field $E_1$, where at increasing temperature the system can exhibit a reentrant behavior from the FE to NC90, back to FE, and again to the NC90 phase at low temperatures. This reentrant behavior, as illustrated in fig.~\ref{fig-e658}, results in a series of jumps of the order parameter and  dielectric permittivity and in barely visible discontinuities of the strain $\eps_3$.  At further increase of temperature the system crosses the supercritical line VIII, at which the permittivity maximum is observed, goes through the ``transitional'' region NC45*, and then undergoes the second order transition again to the FE phase at the line IV, much like at the slightly lower field $E_1=655$~kV/cm, as shown in fig.~\ref{fig-e655}. Again only those phase transitions, where the absolute values of the order parameters are changed, result in discernible jumps of the lattice strain $\eps_3$.

\begin{figure}[hbt]
	\centerline{\includegraphics[width=\columnwidth]{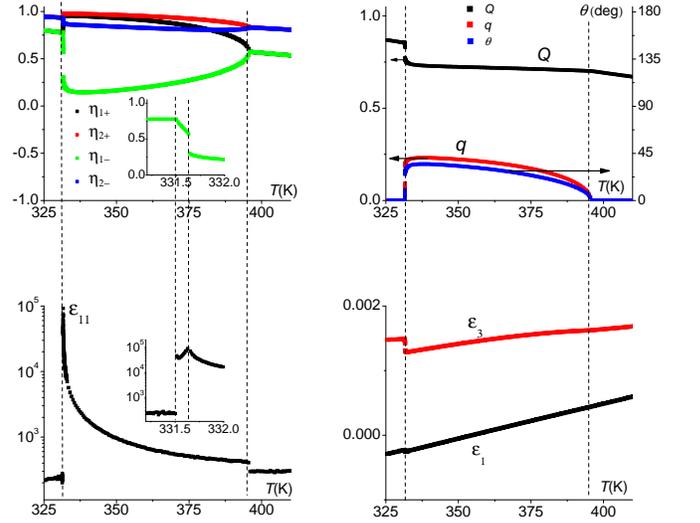}}
	\caption{Same as in fig.~\ref{fig-e100} at $E_1=661.5$~kV/cm. Two vertical dashed lines, going through the two lower phase transitions, overlap. Insets: the temperature dependences of $\eta_{1-}$ and $\eps_{11}$ in the region of these two phase transitions. } \label{fig-e6615}
\end{figure}

At fields between $E_{\textrm{cep}2}$ and $E_{\textrm{cp}3}$ the system is in the FE phase at low to moderate temperatures, then first undergoes the second-order phase transition into the ``transitional'' region NC45* at the line IV and almost immediately the first-order transition at crossing the line III (see fig.~\ref{fig-e6615} and the insets in particular). At higher temperatures the line IV is crossed again, when the system leaves the region NC45* and returns to the FE phase.

\begin{figure}[hbt]
	\centerline{\includegraphics[width=\columnwidth]{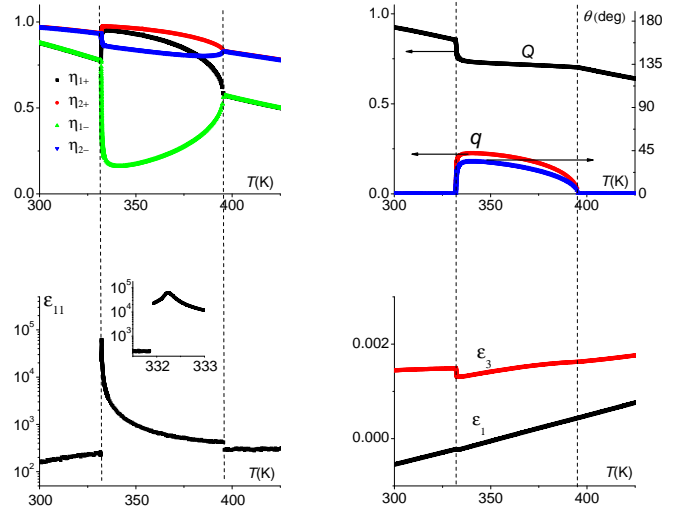}}
	\caption{Same as in fig.~\ref{fig-e100} at $E_1=662$~kV/cm. } \label{fig-e662}
\end{figure}

With increasing field $E_1$ the jumps of the order parameters at the first order transition at the line III decrease, until at  the critical point CP$_3$ the transition becomes of the second order.
At fields just above $E_{\textrm{cp}3}$ (see fig.~\ref{fig-e662}) this transition disappears; instead  only smeared peaks of the permittivity are observed at crossing the supercritical line IX, as seen in the inset to the figure. Apart from that the system behavior is the same as at the fields between $E_{\textrm{cep}2}$ and $E_{\textrm{cp}3}$, depicted in fig.~\ref{fig-e6615}: two second order phase transitions at crossing the line IV twice, with two jumps of the permittivity at the transition points.

\begin{figure}[hbt]
	\centerline{\includegraphics[width=\columnwidth]{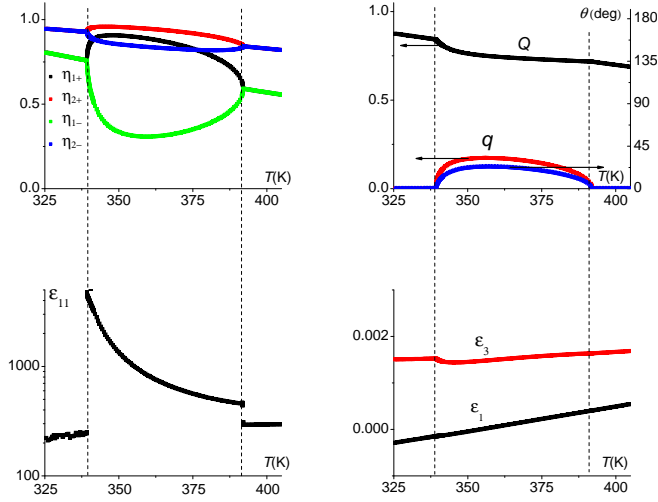}}
	\caption{Same as in fig.~\ref{fig-e100} at $E_1=670$~kV/cm. } \label{fig-e670}
\end{figure}

At higher fields the smeared anomalies, observed earlier at crossing the supercritical line IX, disappear (see fig.~\ref{fig-e670}). The system here still undergoes the two second order phase transitions at the line IV, and the permittivity has jumps at these transitions. At increasing the field these transitions move closer, until at about 690~kV/cm they  merge. Above that field the system is in the FE phase at all temperatures.

The general conclusions that can be drawn about the presented temperature dependences of the order parameters, strains, and the permittivity are the following:

\begin{itemize}
	\item 
The FE ordering parameter $Q$ overall increases with the field at constant temperature, while the AFE order parameter $q$ overall decreases.

	\item
The temperature dependences of the non-collinearity angle $\theta$ closely resemble those of the parameter of $q$, especially at high fields.

\item
The smeared anomalies of the dielectric permittivity (rounded peaks or cusps) at the supercritical lines are accompanied by inflections in the temperature curves of the order parameters $\eta_{f-}$, $Q$, $q$, and of the non-collinearity angle $\theta$.

\item 
The strain $\eps_3$ exhibits discernible discontinuities at those phase transitions, when the absolute values of the order parameters $\eta_{f\pm}$
are changed significantly. At the transitions, when only the sign(s) of
$\eta_{f-}$ are changed (a jump-like sublattice polarization switching)  the strain $\eps_3$ remains unaffected. The same is valid for the strain $\eps_1$. However it has a smaller anomalous part, and the corresponding discontinuities in its temperature curves are even smaller than those of $\eps_3$. The strain $\eps_2$, describing the crystal deformation in the direction perpendicular to the polarization sheets, was equally found to be unaffected by the field $E_1$. These results are in discord with the \textit{ab initio} findings \cite{ishibashi:18}, predicting that the jumps of all three lattice constants are to be observed at the field-induced phase transitions associated with the sublattice polarization rotation, and that the largest discontinuity is to be exhibited by the strain $\eps_1$.
\end{itemize}

\end{document}